\documentclass{emulateapj}

\begin{document}

\title{The origin of dwarf ellipticals in the Virgo cluster}
\author{
A. Boselli\altaffilmark{1}, S. Boissier\altaffilmark{1}, L. Cortese\altaffilmark{2},  
G. Gavazzi\altaffilmark{3}
}
\altaffiltext{1}{Laboratoire d'Astrophysique de Marseille, BP8, Traverse du Siphon, F-13376 Marseille, France}
\altaffiltext{2}{School of Physics and Astronomy, Cardiff University, 5, The Parade, Cardiff CF24 3YB, UK}
\altaffiltext{3}{Universita degli Studi di Milano-Bicocca, Piazza delle Scienze 3, 20126
Milano, Italy}

\begin{abstract}

We study the evolution of dwarf ($L_H$ $<$ 10$^{9.6}$ L$_{H \odot}$) 
star forming and quiescent galaxies in the Virgo
cluster by comparing their UV to radio centimetric properties to the
predictions of multizone chemo-spectrophotometric models of galaxy
evolution especially tuned to take into account the perturbations
induced by the interaction with the cluster intergalactic medium. 
Our models simulate one or
multiple ram pressure stripping events and galaxy starvation. Models predict that
all star forming dwarf galaxies entering the cluster for the first
time loose most, if not all, of their atomic gas content, quenching on
short time scales ($\leq$ 150 Myr) their activity of star
formation. These dwarf galaxies soon become red and quiescent,
gas metal-rich objects with spectrophotometric and structural properties similar to those of
dwarf ellipticals. Young, low luminosity, high surface brightness star forming galaxies
such as late-type spirals and BCDs are probably the progenitors of relatively massive dwarf ellipticals,
while it is likely that low surface brightness magellanic irregulars evolve into 
very low surface brightness quiescent objects hardly detectable in ground based imaging surveys.
The small number of dwarf galaxies
with physical properties intermediate between those of star forming
and quiescent systems is consistent with a rapid ($<$ 1 Gyr) transitional phase between
the two dwarf galaxies populations.
These results, combined with statistical considerations, are
consistent with the idea that most of the dwarf ellipticals dominating
the faint end of the Virgo luminosity function were initially star forming
systems, accreted by the cluster and stripped
of their gas by one or subsequent ram pressure stripping events.
\end{abstract}


\setcounter{footnote}{0} 

\section{Introduction}

Dwarf galaxies ($M_B>$-18) are the most common objects in the nearby
universe (Ferguson \& Binggeli 1994). Their importance resides in the
fact that they represent in cold dark matter models the building blocks
of hierarchical galaxy formation (e.g. White \& Rees 1978, White
\& Frenk 1991). Their study is thus fundamental for constraining
models of galaxy formation and evolution.\\ 
Observations of dwarf
galaxies, necessarily limited to the nearby universe, revealed however a
more complex origin than that predicted by models, making this
class of galaxies even more interesting. Among dwarf galaxies, dwarf
ellipticals (dE) and the less luminous dwarf spheroidals
(dSph)\footnote{Unless specified, in the following we indicate with dE
all quiescent dwarf galaxies, including dSph} are more common than
star forming Im and BCDs (Ferguson \& Binggeli 1994). Firstly thought
as the low-luminosity extension of bright ellipticals, several observational
evidences indicate that they form an independent class of
objects (Bender et al. 1992). Dwarf and giant ellipticals have
Sersic light profiles of index $1/n$, with $n$ progressively 
increasing with luminosity (Graham \& Guzman 2003; 
Gavazzi et al. 2005a). The colour magnitude relations
measured using resolved stars in local group dwarfs spheroidals 
(Mateo 1998; Grebel 1999), the optical (Conselice et al. 2003a) and UV
(Boselli et al. 2005a) integrated color magnitude relations, the
subsolar [$\alpha$/Fe] ratios (Van Zee et al. 2004a) and
the UV to near-IR spectral energy distribution of dE in the Virgo
cluster (Gavazzi et al. 2002a), however, all indicate a more gradual star
formation history with recent episodes of activity in these
low-luminosity quiescent systems than in massive ellipticals. This constitutes
the first evidence against their very old origin. A further
disagreement with model predictions is that, although more frequent
than bright galaxies, their number density is still significantly smaller than 
predicted by hierarchical models of galaxy formation for the field
luminosity function (Kauffmann et al. 1993; Cole et al. 1994;
Somerville \& Primack 1999, Nagashima et al. 2005) or for the local
group (Klypin et al. 1999; Bullock et al. 2000).\\
The strong similarities in the structural properties of star forming and
quiescent dwarf galaxies in the nearby universe, namely their similar
optical morphology and light profiles (both roughly exponentials), suggested
that dwarf ellipticals might result from gas removal and subsequent
suppression of star formation in gas rich dwarf galaxies. 
Gas removal might result either from its blowout due to the
kinetic energy injected into the ISM by supernova explosion following
a strong burst of star formation (Dekel \& Silk 1986, Vader 1986,
Yoshii \& Arimoto 1987), gas exhaustion through subsequent episodes
of star formation (Davies \& Phillipps 1988), or from external
perturbations induced by the hostile environment in which galaxies
evolve.  External perturbations include tidally
induced mass loss in high speed encounters (Moore et al. 1998), tidal
stirring (Mayer et al. 2001a, 2001b) and ram-pressure stripping
induced by nearby companions (Lin \& Faber 1983) or by the hot
intergalactic medium in massive clusters (van Zee et al. 2004b). \\
Several observational evidences favor the environmental
scenario against the gas blowout due to supernova explosions. 
The clearest indication that the environment plays an
important role in the formation and evolution of dwarfs is the
morphology segregation effect (Dressler 1980) which extends to
low-luminosity systems (Binggeli et al. 1988; 1990; Ferguson \&
Binggeli 1994). Furthermore it has been recently claimed that the
removal of the ISM through supernova winds is quite difficult in
low-luminosity, dark matter dominated systems (Mac Low \& Ferrara
1999; Ferrara \& Tolstoy 2000; Silich \& Tenorio-Tagle 1998, 2001).
The discrete star formation history of several dwarf galaxies in the
local group, as deduced by the analysis of their colour magnitude
relation (Mateo 1998; Grebel 1999), is a further indication that dwarf
spheroidals can retain their ISM through several episodes of star
formation.\\
Another observational evidence favoring the transformation
of star forming galaxies into quiescent dwarf ellipticals is the
presence of rotationally supported (Pedraz et al. 2002; Geha et
al. 2003; van Zee et al. 2004b) and/or HI gas rich (Conselice et
al. 2003b; van Zee et al. 2004b) dE in the Virgo cluster. Recent
studies based on SDSS imaging and spectroscopic data have shown that
$\sim$ 50 \% of the bright end of the dE galaxy population in the
Virgo cluster is characterized by disk features such as spiral arms or
bars, this fraction decreasing down to $\sim$ 5\% at lower
luminosities (Lisker et al. 2006a; see also Graham et al. 2003). Meanwhile $\sim$ 15\% of
the bright dE in Virgo have blue centers revealing a recent 
activity of star formation. From a statistical point of view, the
line-of-sight velocity distribution of dE inside clusters is similar to that of
late-type galaxies suggesting a recent infall (Binggeli et al. 1993;
Conselice et al. 2001).\\ 
Not all observational
evidences, however, are consistent with the transformation of 
dwarf star forming galaxies
into dwarf ellipticals under the effect of the environment. To reproduce the
color magnitude relation of dwarf ellipticals, it has been claimed that 
magellanic irregulars
should have faded $\sim$ 1.5 magnitudes in the B band thus reaching
surface brightnesses weaker than $\mu(B)_e$ $=$ 25 mag arcsec$^{-2}$,
values significantly smaller than the observed ones (Bothun et
al. 1986). Most of the bright dwarf ellipticals in the Virgo cluster
have a bright nucleus (Binggeli et al. 1985; Ferguson \&
Binggeli 1994) of small size, as shown by HST observations ($\sim$ 4 pc, C\^ote et al. 2006) 
while dwarf irregulars do not. While both dwarf
irregulars and dwarf ellipticals follow different
metallicity-luminosity relations, dwarf spheroidals are more
metal-rich than dwarf irregulars at the same optical luminosity
(Grebel et al. 2003 and references therein). The flattening
distribution of non nucleated dE is similar to that of late-type spirals, Im
and BCD (Binggeli \& Popescu 1995). This class of objects, however, is significantly less round than nucleated systems. 
Although they exist, the fraction of objects belonging to the
intermediate dIrr/dE transition class is too small. Furthermore a
simple transformation of Im galaxies recently infalling in the cluster
into dE does not seem to reproduce the observed difference in the cluster and
field luminosity functions (Conselice 2002). Other strong constraints
against the transformation of star forming into quiescent dwarfs are
given by the studies of globular clusters, taken as a probe of the
early phases of galaxy formation. The specific frequency of globular
clusters in dwarf ellipticals, in fact, is significantly higher than
that of star forming galaxies. This result has been interpreted 
as an evidence for a different formation scenario for the two 
galaxy populations (Miller et al. 1998; Strader et
al. 2006).\\ 
With the aim of explaining these evidences, slightly
different evolutionary scenarios have been proposed: cluster dwarf
ellipticals might have been formed from higher surface brightness BCD
galaxies (Bothun et al. 1986), or from tidally induced mass loss in
multiple high speed encounters of massive galaxies (galaxy harassment;
Mastropietro et al. 2005). Alternatively, Lisker et al. (2006a; 2006b;
2007) proposed that the cluster dwarf elliptical galaxy population is
composed of different subcategories of objects, of which not all have
been formed from gas stripped star forming dwarfs. As here emphasized 
the issue is still hotly debated.\\ 
A few years ago we started collecting multifrequency data covering the
whole electromagnetic spectrum for galaxies in nearby clusters in
order to study the effects of the environment on galaxy evolution. 
Up to now our
research was primarily focused on the bright end of the luminosity
function. Our interest covered the present and past star formation
activity (Gavazzi et al. 1991, 1998, 2002b, 2006a), the atomic
(Gavazzi et al. 2005b, 2006b) and molecular gas content (Boselli 1994;
Boselli et al. 1994, 1997a, 2002), the radio continuum and IR
properties of cluster galaxies (Gavazzi \& Boselli 1999; Gavazzi et
al. 1991). The results of our analysis, combined with
those obtained by other teams, are summarized
in a recent review article (Boselli \& Gavazzi 2006).\\ 
The aim of the
present paper is to extend the multifrequency analysis 
to the low-luminosity end of the luminosity function with the
purpose of studying the possible transformation of star forming
objects into dwarf ellipticals. This exercise is here done using a
complete sample of galaxies in the Virgo cluster. The novelty of this
work compared to previous investigations is twofold: the analyzed
sample has an unprecedented multifrequency spectral coverage from the UV to near-IR
imaging data. Furthermore this unique
dataset is compared to the predictions of multizone
chemo-spectrophotometric models of galaxy evolution here adapted to
take into account the perturbations induced by the cluster
environment. The type of perturbation simulated by the models are
ram-pressure stripping (Gunn \& Gott 1972) and starvation (Larson et
al. 1980).  The combination of multifrequency data with these models
on the galaxies NGC 4569 (Boselli et al. 2006) and NGC 4438 (Boselli
et al. 2005b) indeed indicated how powerful this method is for
studying and constraining the evolution of cluster galaxies. \\
Our previous investigations have
emphasized that the most important parameter governing the
evolution is the total mass as traced by the near-IR H band luminosity
(Gavazzi et al. 1996; 2002a; Boselli et al. 2001). Below a 
certain mass ($L_H$ $<$ 10$^{9.6}$ L$_{H \odot}$)
we identify the sequence of dwarf galaxies 
that we subdivide into quiescent and star forming disregarding
their detailed morphology.\\
A major uncertainty in the interpretation of the
multifrequency data, and in particular of those at short wavelength,
is the extinction correction, which however is expected to be minor in
low-metallicity star forming dwarf systems (Buat \& Xu 1996) and
probably negligible in dwarf ellipticals. For these reasons we decided
to limit the present analysis to the cluster dwarf galaxy population,
leaving the discussion relative to massive galaxies to a future
communication. With the aim of driving the reader's attention 
to the scientific results of this work, the presentation of the dataset
and the general description of the models are given in Appendix.

\section{Sample and data}

The analysis presented in this work is based on an optically selected
sample of galaxies in the Virgo cluster (12h $<$ R.A.(2000) $<$13h;
0$^o$ $<$ dec $<$ 18$^o$) extracted from the Virgo Cluster Catalogue
(VCC) of Binggeli et al. (1985), with $m_B$ $<$18
mag (that for a distance of 17 Mpc gives $M_B$ $\le$ -13.15) which corresponds to its completeness limit.\\ 
Although the present analysis is focused on the
dwarf star forming (Scd-Im-BCD) and quiescent (dE-dS0) galaxy
populations, that we define as $L_H$ $<$ 10$^{9.6}$ L$_{H \odot}$
(which roughly corresponds to $M_B$ $\ga$ -18 mag),
we also include brighter objects selecting all galaxies with
$m_B$ $<$ 18 mag. This are all bona-fide Virgo
cluster members, whose distances have been assigned following the
subcluster membership criteria of Gavazzi et al$.$ (1999) which is based on 
combined position and redshift data (now available for 83\% of the objects). 
The adopted distance for each subgroup is: 17 Mpc for cluster A, 
the North and East clouds and the Southern extension, 23 Mpc
for cluster B and 32 Mpc for the W and M clouds. The selected sample 
includes a total of 1010 galaxies, 445 of which are classified as dE or
dS0, 215 as Scd-Sd, Im or BCD and 45 as dE/Im or ? in the VCC.
Because of their intrinsic color, 
the lowest H band luminosities 
(thus the lowest total dynamical masses) are reached only in star forming dwarfs.\\
Model predictions are compared to UV to near-IR imaging and spectroscopic data collected thanks to our own 
observations (optical, near-IR and H$\alpha$ imaging, optical integrated spectroscopy)
or from different ground based (SDSS) and space (GALEX) missions as well as from multifrequency data 
available in the literature. The detailed description of the
used dataset is given in the Appendix.

\section{The evolution of low-luminosity late-type galaxies in clusters: model predictions}

The evolution of galaxies is traced using the multi-zone chemical and spectro-photometric
models of Boissier \& Prantzos (2000), updated with an empirically-determined star
formation law (Boissier et al$.$ 2003) relating the star formation
rate to the total-gas surface densities, and modified to simulate the effects induced 
by the interaction with the cluster IGM. In the starvation scenario (Larson et al. 1980, Balogh et al. 2000,
Treu et al. 2003), the cluster acts on large scales by removing any
extended gaseous halo surrounding the galaxy, preventing further infall
of such gas onto the disk. The galaxy then become anemic simply because it 
exhausts the gas reservoir through on-going star formation. Starvation has been simulated 
by stopping infall in the model.\\ 
The ISM of galaxies crossing the cluster with velocities of $\sim$ 1000 km s$^{-1}$ can be removed by 
the ram pressure exerted by the hot and dense IGM (Gunn \& Gott 1972). Gas removal induces a quenching of the
star formation activity (Boselli \& Gavazzi 2006).
The ram pressure stripping event is simulated by assuming a gas-loss rate inversely proportional to the potential of the galaxy,
with an efficiency depending on the IGM gas density radial profile of the Virgo cluster given by Vollmer et al. (2001).
The details of the models are given in the Appendix. Here 
we present their results.

\subsection{The starvation scenario}

\begin{figure*}
\includegraphics[width=13cm,angle=0]{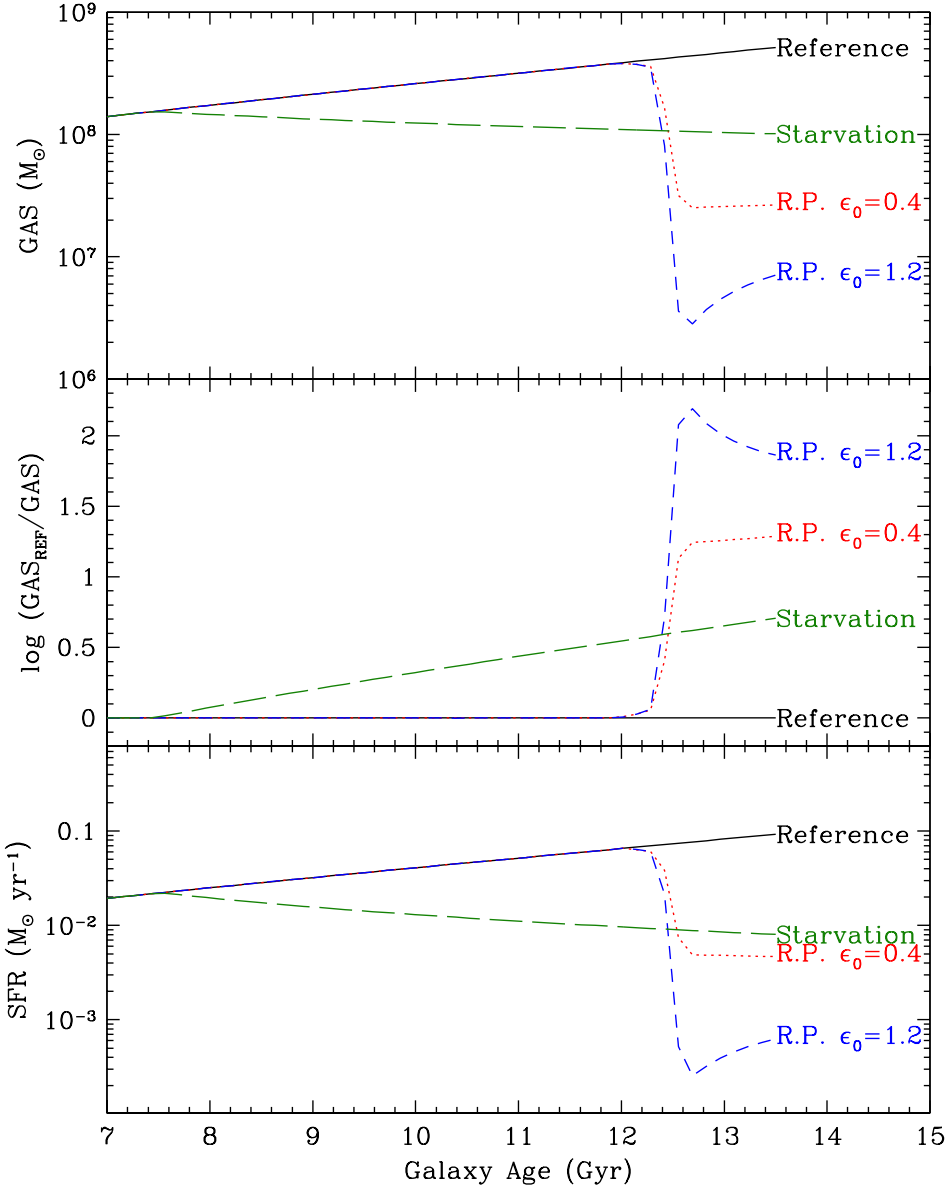}
\small{\caption{The evolution with time of the total gas content (upper panel), of the
unperturbed to perturbed gas mass ratio (central panel, in logarithmic scale: 
this entity corresponds to the HI-deficiency parameter) and of 
the star formation rate (lower panel) for a galaxy with rotational velocity $V_C$ = 55 km s$^{-1}$
and spin parameter $\lambda$=0.05. The continuum black line shows the unperturbed model (reference),
the green long dashed line a starvation model for a galaxy-cluster interaction started 6 Gyr ago, 
the red dotted line a ram pressure stripping model with an efficiency of 
$\epsilon_0$ = 0.4 M$\odot$ kpc$^{-2}$ yr$^{-1}$, the blue short dashed line
a ram pressure stripping model with an efficiency of $\epsilon_0$ = 1.2 M$\odot$ kpc$^{-2}$ yr$^{-1}$. 
The ram pressure stripping event in both of these models started 1 Gyr ago ($t_{rp}$=1 Gyr). 
The increase with time of the total gas content and of the star formation activity for an unperturbed galaxy 
is due to gas infall. 
\label{figure1}}}
\end{figure*}

\begin{figure*}
\epsscale{1.0} \includegraphics[width=15cm,angle=0]{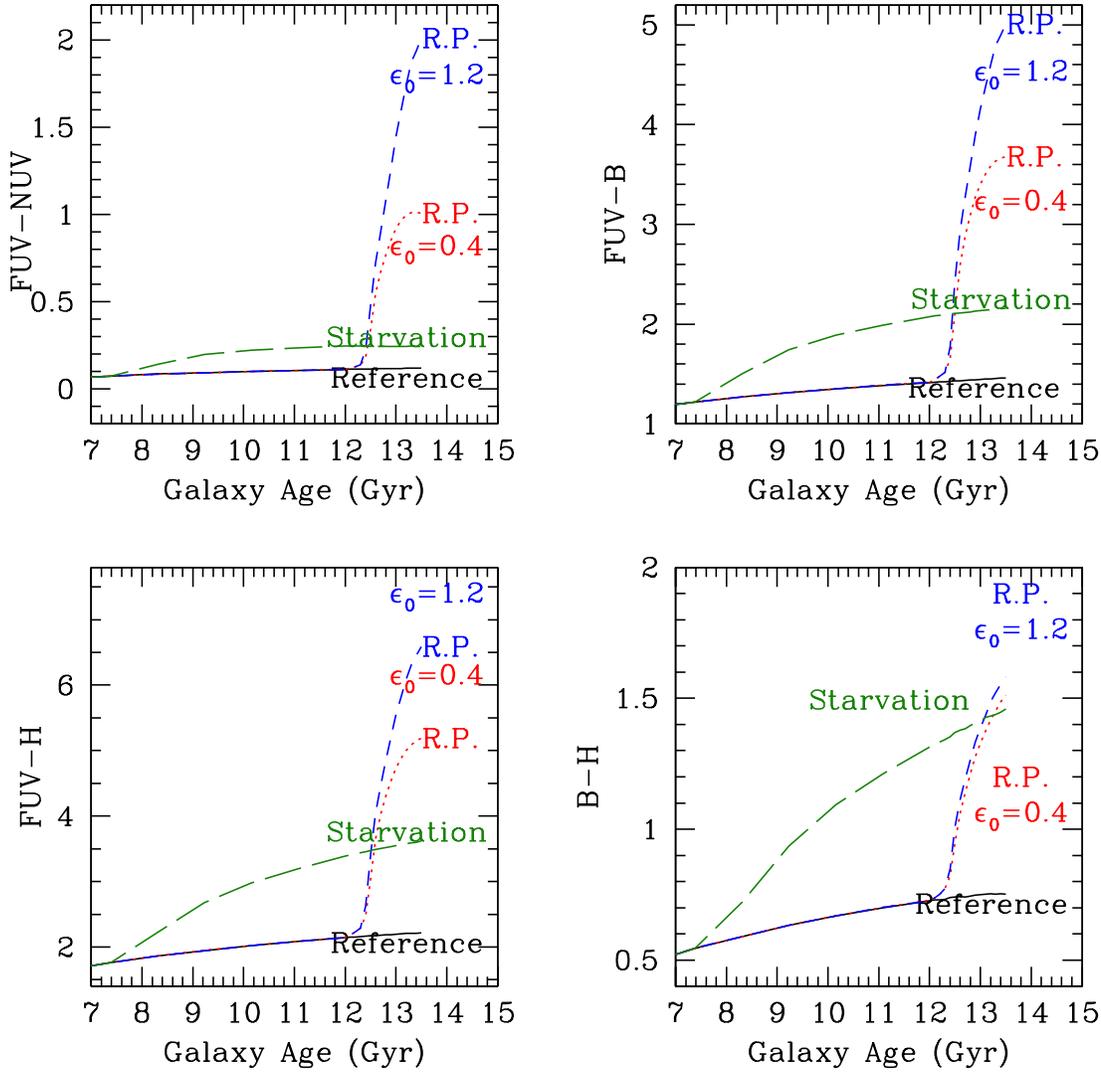}
\small{\caption{The evolution of different color indices (in the AB system) for a galaxy with rotational velocity $V_C$ = 55 km s$^{-1}$
and spin parameter $\lambda$=0.05. Symbols are as in Fig. \ref{figure1}.
\label{figure2}}}
\end{figure*}

\begin{figure}
\epsscale{1.0} \includegraphics[width=15cm,angle=0]{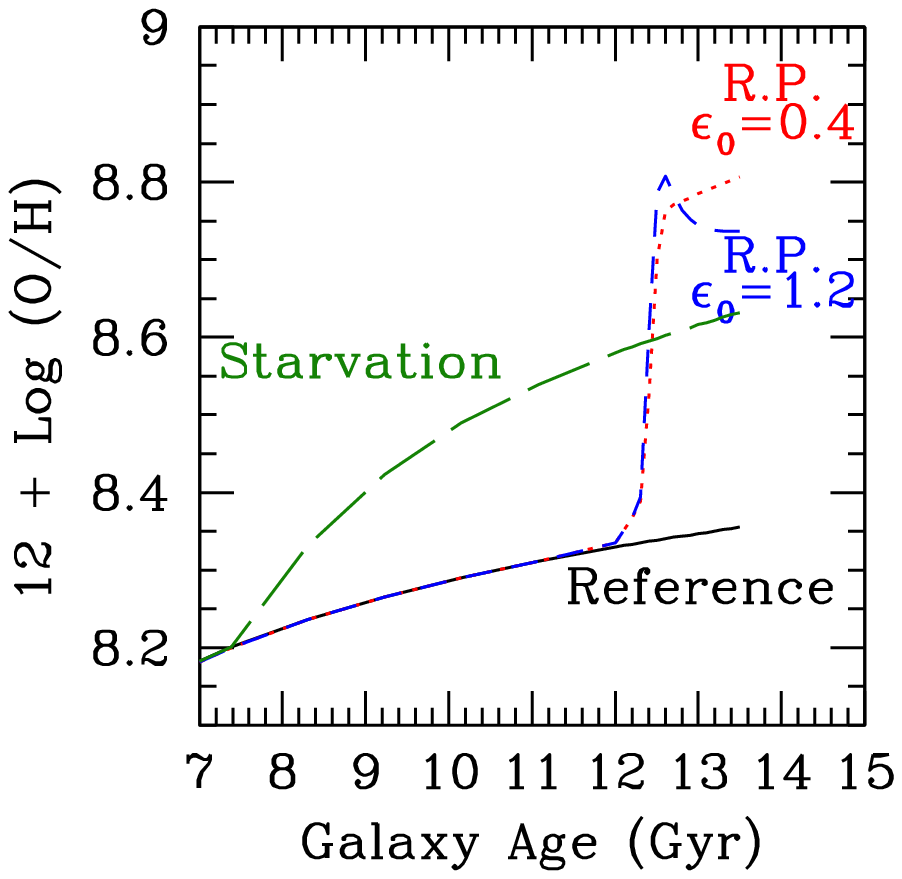}
\small{\caption{The evolution of the gas 12 $+$ Log(O/H) 
metallicity index 
for a galaxy with rotational velocity $V_C$ = 55 km s$^{-1}$
and spin parameter $\lambda$=0.05. Symbols are as in Fig. \ref{figure1}.
\label{figure4}}}
\end{figure}

The total gas content, the activity of star formation and the
metallicity of dwarf galaxies are barely affected by starvation unless
the interaction started long time agoo, as shown for a model galaxy of rotational velocity
$V_C$=55 km s$^{-1}$ and spin parameter $\lambda$ = 0.05 (Fig. \ref{figure1} to \ref{figure5}). 
HI-deficiencies of
$\sim$ 0.8, similar to the average value of Virgo galaxies within the
virial radius of the cluster, can be obtained only if starvation
started more than 6 Gyr ago (see Fig. \ref{figure1}). \\
Gas removal induces a mild decrease of the star formation activity (Fig. \ref{figure1}).
For this reason galaxies suffering starvation for a 
long time are redder than similar unperturbed objects, in particular in
those color indices tracing significantly different stellar populations 
($FUV-B$, $FUV-H$, $B-H$; Fig. \ref{figure2}). 
The $FUV-NUV$ color index is barely affected, because 
starvation acts on longer timescales than the lifetimes of the stellar 
populations emitting in the two UV filters ($\leq$ 10$^8$ years)\footnote{The UV emission of 
star forming galaxies is due to relatively young and massive A stars (Boselli et al. 2001). 
The UV emission of dwarf ellipticals is due to a residual star formation activity
since does not show, as in massive ellipticals, the UV upturn due to hot, 
evolved stars (Boselli et al. 2005a).}, and does not
totally eliminate these populations as significant amount of star formation, although
reduced, still takes place. 
%
\\
Gas metallicity increases 
more than in the unperturbed case since newly formed 
metals are diluted in smaller amounts of gas, and since the
metallicity is not reduced by the infall of pristine gas (Fig. \ref{figure4}).\\
The effects of starvation on the structural properties of galaxies 
(provided that the interaction started a long time ago)
are major.
%
The decrease of the surface brightness is expected to be $\sim$ 1
mag in the H band 
to $\sim$ 3 mag in the FUV band 
for an interaction started 6 Gyr ago (Fig. \ref{figure5}).
The effect would be significantly smaller for a recent interaction. \\
The effects of starvation 
on the effective radius are minor since starvation is a global effect, and thus does 
not depend on radius. Nevertheless, an effect on the effective
radius is visible, especially at short wavelengths. The reason is the
radial dependence of infall (galaxies are formed inside-out in our
model). When the starvation starts, infall has occurred in the inner
zones, where relatively large amount of gas can be found (and thus
high level of star formation and UV surface brightness). This gas is
progressively depleted by star formation, leading to a decrease of the
inner surface brightness while the outer parts, where the
star formation is low because of the low amount of gas present, 
dim at a lower rate. The net effect is an increase
in the effective radius.

\subsection{The ram pressure stripping scenario}

\begin{figure*}
\epsscale{1.0} \includegraphics[width=15cm,angle=0]{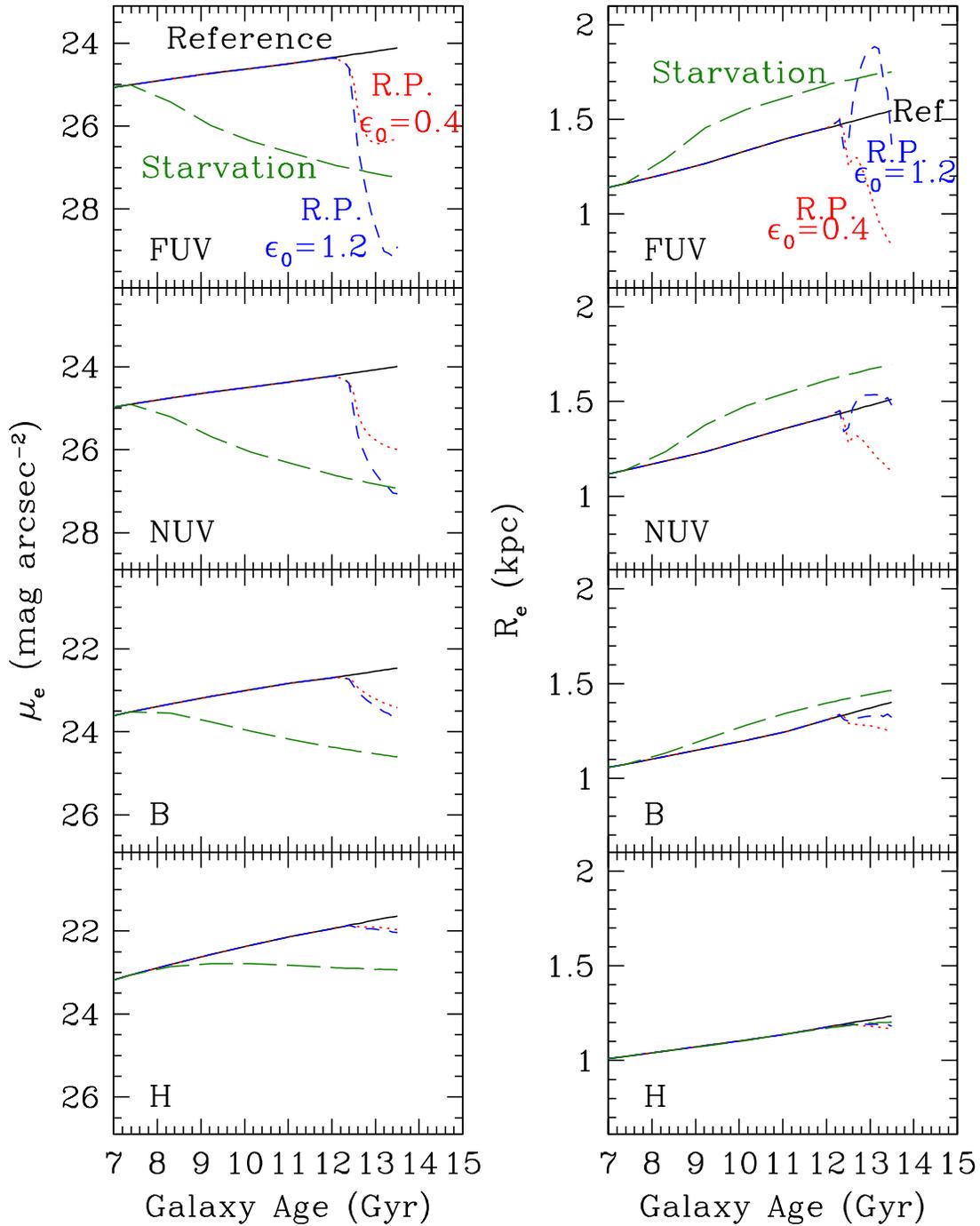}
\small{\caption{The evolution of the effective surface brightness (left panels)
and radius (right panels) in different bands
for a galaxy with rotational velocity $V_C$ = 55 km s$^{-1}$
and spin parameter $\lambda$=0.05. Symbols are as in Fig. \ref{figure1}.
\label{figure5}}}
\end{figure*}

\begin{figure*}
\epsscale{1.0} \includegraphics[width=15cm,angle=0]{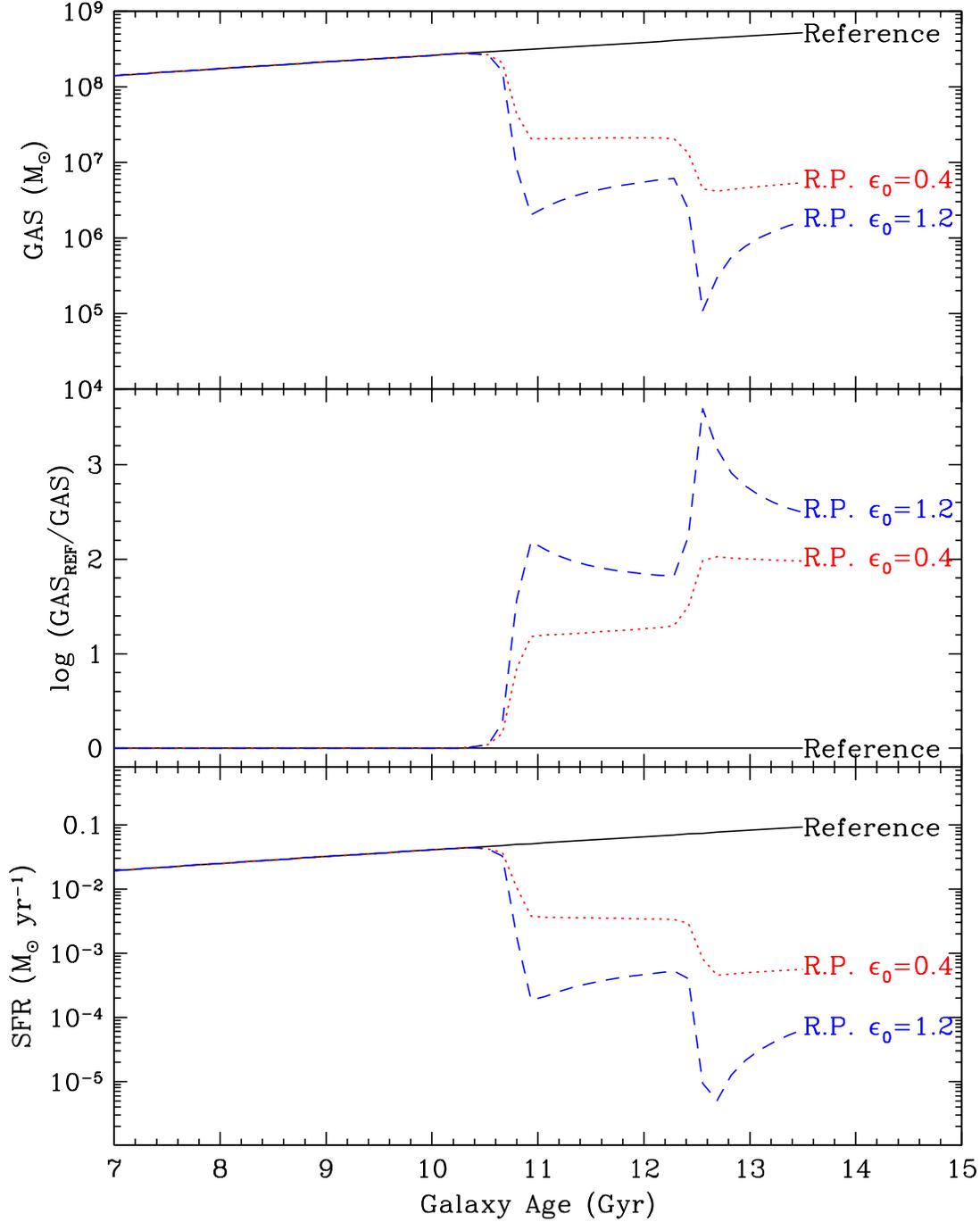}
\small{\caption{The evolution with time of the total gas content (upper panel), of the
unperturbed to perturbed gas mass ratio (central panel) and of 
the star formation rate (lower panel) for a galaxy with rotational velocity $V_C$ = 55 km s$^{-1}$
and spin parameter $\lambda$=0.05 after two ram pressure stripping events. 
The continuum black line shows the unperturbed model (reference), 
the red dotted line a ram pressure stripping model with an efficiency of 
$\epsilon_0$ = 0.4 M$\odot$ kpc$^{-2}$ yr$^{-1}$, the blue short dashed line
a ram pressure stripping model with an efficiency of $\epsilon_0$ = 1.2 M$\odot$ kpc$^{-2}$ yr$^{-1}$. 
The two stripping events are delayed by 1.7 Gyr, the crossing time of late type galaxies in the Virgo cluster.
\label{multiple}}}
\end{figure*}

The effects of a ram-pressure stripping event started 1 Gyr ago ($t_{rp}$=1 Gyr) on the evolution of total gas
content, star formation activity, stellar populations (UV to near-IR
color indices), metallicity and
structural (effective radius and surface brightness) parameters of
dwarf galaxies as predicted by our models are also shown in
Fig. \ref{figure1} to \ref{figure5}.
Because of the shallow potential
well of dwarf galaxies, most of their gas is efficiently removed on
very short time scales ($\sim$ 150 Myr), leading to the formation of
extremely gas deficient objects with HI-deficiencies $\sim$ 2 for
$\epsilon_0$ = 1.2 M$\odot$ kpc$^{-2}$ yr$^{-1}$, or $\sim$ 1.2 for
$\epsilon_0$ = 0.4 M$\odot$ kpc$^{-2}$ yr$^{-1}$ ($\epsilon_0$ is 
the parameter that quantifies the efficiency of the process of 
ram pressure stripping; see the Appendix B).\\
As a consequence of this lack of gas, star formation abruptly
decreases by a factor of $\sim$ 10 to $\sim$ 100 and continues at a
very low level afterward. In the $\epsilon_0$ = 1.2 M$\odot$
kpc$^{-2}$ yr$^{-1}$ model, the gas content progressively increases
after the interaction as stars of small masses reach their end of
life point and return more gas to the interstellar medium (we do
not use the instantaneous recycling approximation). The star
formation rate is fed by this recycled gas, and increases as
well, staying nevertheless at low levels. In models with a lower ram
pressure efficiency, the recycled gas represents a small amount
with respect to the one left after
the interaction. As a result, this effect is not visible: the gas and the
star formation rate are about constant after the event.\\
Stopping the star formation makes galaxies redder. Since the decrease of the star formation activity is
rapid, the younger stellar populations, those emitting in the FUV and NUV filters, 
are the most affected: the $FUV-H$ color index, for instance, increases by $\sim$  4.5 magnitudes
in 1 Gyr, while $B-H$ by only 1 magnitude. \\
%
As for the starvation scenario, the gas metallicity
index significantly increases with time. The increase of the gas
metallicity is higher during a ram pressure stripping event since 
virtually all the gas has been removed, while some gas is left in the disk 
during a starvation event. There is thus less dilution for a ram pressure
event than in the occurrence of starvation.\\
Stopping the star formation drastically changes the structural properties of galaxies. 
Stripped galaxies have lower surface brightnesses and are slightly smaller with respect to
unperturbed objects when observed at short wavelengths. At long wavelengths, where the emission 
is dominated by the evolved stellar populations, the effects are almost negligible (Fig. \ref{figure5})
\footnote{In the UV, in our example of Fig.  \ref{figure5}, the
  effective radius presents large variations in its value after ram
  pressure started. This is due to the fact that the galaxy is left
  with very low amount of gas, and the effective radius becomes
  sensitive to two opposite effects. In a first phase, the dominant
  effect is that the inner disk is depleted through star formation at
  a larger rate than the outer disk, leading to an increase of $R_{e}$
  (like in the starvation case). In a second phase, the inner disk is
  partially replenished with recycled gas (ejected at the end of the
  life of long-lived stars), leading this time to a decrease of
  $R_{e}$.}.
%
%
%
\\
An accurate comparison between the effects induced by ram-pressure
stripping and starvation on dwarf galaxy properties indicates that the
former interaction is more efficient than the latter in removing gas
and suppressing any star formation activity in galaxies on short time
scales. To be efficient, starvation should have started several Gyr
ago. Recent ram pressure stripping events, however, did not have
enough time to perturb the old stellar component emitting in the
near-IR. 
%
\\
The effects of multiple encounters on the gas content and star
formation activity of a dwarf, star forming galaxy of spin parameter
$\lambda$ = 0.05 and rotational velocity $V_C$ = 55 km s$^{-1}$ are
shown in Fig. \ref{multiple}.
The comparison of Fig. \ref{multiple} with Fig. \ref{figure1} (single
stripping event) shows how multiple interactions make galaxies even
poorer of gas and, as a consequence, redder in colors, than similar
objects that underwent a single ram pressure stripping event. 
For low efficiencies, successive
crossings are almost additive (some gas left on the first crossing is
removed on the next one). For large efficiencies, the second crossing, occurring
1.7 Gyr after the first one, mostly remove any recycled gas, quenching a possible
increase of star formation. In both cases, multiple crossing galaxies
end up as redder, poorer in gas objects. A summary of the predicted
perturbations induced by ram pressure stripping and starvation on the physical and structural 
properties of dwarf galaxies are given in Table \ref{Tabmod}.

\section{Models vs. observations}

The multifrequency dataset available for dwarf galaxies in the Virgo cluster (see Appendix) allows us to compare model predictions
with observations. 

\subsection{Gas content}

\begin{figure*}
\epsscale{1.0} \includegraphics[width=23cm,angle=0]{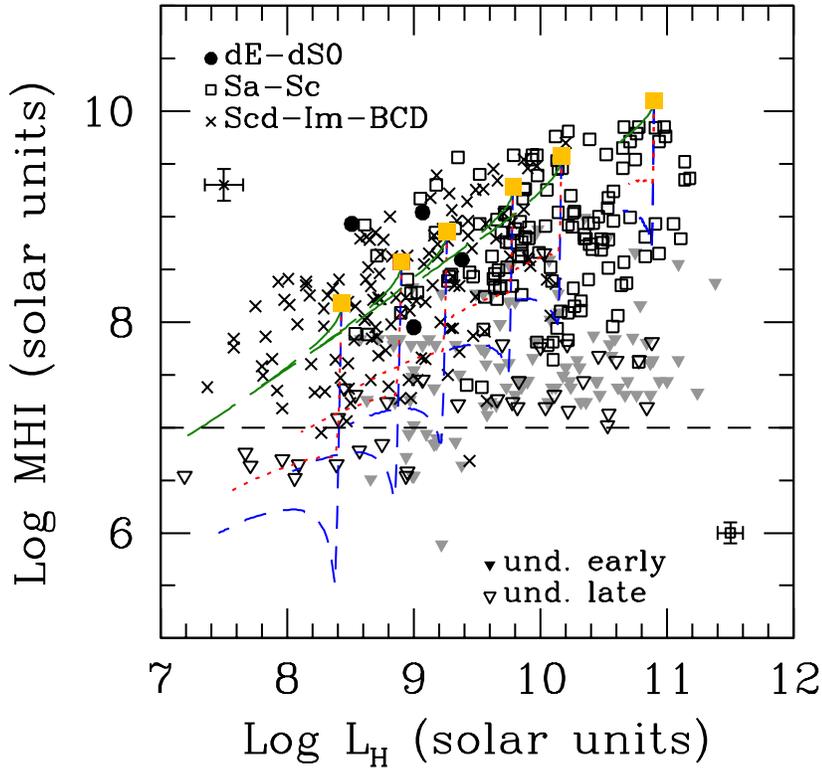}
\small{\caption{The relationship between the HI mass and the H band luminosity, tracer of the total mass of
galaxies, compared to model predictions for different lookback times of
a ram pressure/starvation event. Empty squares are for Sa-Sc
galaxies, crosses for Scd-Im-BCD and filled
circles for dE-dS0. Open triangles are upper limits for undetected late-type
galaxies, gray filled triangles for undetected early-types. The dashed
line indicates the region which is inaccessible to the observations
with a sensitivity of 1 mJy per channel for galaxies at the distance
of Virgo. Model predictions for unperturbed galaxies of spin parameter
$\lambda$ = 0.05 and $V_C$ = 40, 55, 70, 100, 130 and 220 km s$^{-1}$
are indicated with orange squares (from left to right). Model
predictions for a ram-pressure stripping event at different epochs are
indicated by blue dashed ($\epsilon_0$ = 1.2 M$\odot$ kpc$^{-2}$
yr$^{-1}$) and red dotted ($\epsilon_0$ = 0.4 M$\odot$ kpc$^{-2}$
yr$^{-1}$) lines, while the starvation scenario by the long-dashed
green line. 
One should take care that the models presented in this figure and later
are not evolutionary tracks, but result from the combined effects of ram
pressure, recycled gas and epoch of the interaction. Especially, we note that
for interaction that occurred a long time ago, the available gas at that time
was smaller (the galaxy was not yet fully formed), and as a result, the final gas
amount is also smaller.
\label{def}}}
\end{figure*}

Ram-pressure stripping models are able to reproduce the HI-deficient\footnote{Model HI gas masses are estimated assuming that the total
gas amount of star forming galaxies is composed of $\sim$ 15\% of molecular gas (Boselli et al. 2002)
and 30\% of helium, thus $MHI$ $\sim$ $M_{gas}$/1.4.} galaxy
population while starvation does not (Fig. \ref{def}) \footnote{We notice that here and in the subsequent figures 8 to 10 and 12
the model tracks do not describe a time evolution, but rather
show current values for models with different lookback times of a ram
pressure or starvation event.}.
As previously remarked in sect. 3.2, ram-pressure stripping can produce 
HI-deficient massive galaxies ($L_H$ $\sim$ 10$^{11}$ L$_{H \odot}$) such as those observed in Virgo. 
For a massive galaxy with $V_C$ = 220 km s$^{-1}$ and spin parameter $\lambda$=0.05, a ram pressure stripping event started 1 Gyr ago
with $\epsilon_0$ = 0.4 M$\odot$ kpc$^{-2}$ yr$^{-1}$ leads to an HI-deficiency of $\sim$ 0.8, similar to the average HI-deficiency of
massive spiral galaxies in the center of the Virgo cluster (Gavazzi et al 2005b; Boselli \& Gavazzi 2006).
More extreme HI-deficient objects of the cluster (HI-deficiency $\sim$ 1.4) 
can be produced by assuming $\epsilon_0$ = 1.2 M$\odot$ kpc$^{-2}$ yr$^{-1}$.
NGC 4569 can be taken as a typical object of this second category (Boselli et al. 2006).\\
We notice however that the HI emission of the ram-pressure stripped dwarf galaxy population
can be hardly detected by the recent HI surveys such as ALFALFA (Giovanelli et al. 2005)
and AGES (Auld et al. 2006), able to reach only galaxies with HI gas masses  
$\geq$ 10$^{7.5}$ M$\odot$ (ALFALFA) or slightly smaller (AGES) at the distance of Virgo (17 Mpc)
(dashed line in Fig. \ref{def}). 
Galaxies less luminous than $L_H$ $\sim$ 10$^{9.2}$ L$_{H \odot}$ would be detected only in the case 
stripping is still under way. Since the gas stripping event is short in time ($\sim$ 150 Myr), 
we expect that these objects are extremely rare.

\subsection{Stellar populations}

Figure \ref{colmag} shows several color-magnitude relations for
galaxies in the Virgo cluster. We easily recognize the well known red
and blue sequences relative to the quiescent and star forming galaxy
populations (Scodeggio et al. 2002; Gil de Paz et al. 2007; see also
Visvanathan \& Sandage 1977; Bower et al. 1992 for early-type galaxies
and Tully et al. 1982, Gavazzi et al. 1996 for late-type galaxies)
and a relatively large number of objects in between these two sequences (see
Sec. 5). We
remind that the reddest quiescent dwarfs might be undetected in the UV
bands because of the sensitivity of GALEX. Model predictions clearly
show that only a ram-pressure stripping event (and not starvation) 
is able to significantly 
modify the colors of galaxies. As expected the effects are more
important at short than at long wavelengths, affecting more colors
as e.g. $FUV-H$, $FUV-B$ than e.g. $B-H$. The colour index of bright spirals ($L_H$ $\geq$
10$^{10}$ L$_{H \odot}$) can increase by up to $\sim$ 1 mag in
$FUV-NUV$ and 2 mag in $FUV-H$ or $FUV-B$ $\sim$ 500 Myr after the
interaction, making galaxies as red as the most deficient spirals
located in between the red and blue sequences in the color-magnitude
diagram. NGC 4569 might be one of these objects.  \\
The effect on dwarfs is more important, making star forming
irregulars as red as dwarf ellipticals of similar luminosity in the
case of an efficient gas stripping event or just after they crossed
twice the cluster core ($\sim$ 2 Gyr), as shown in
Fig. \ref{colmag_2} (see also Fig. \ref{sed_ref}).\\

\subsection{Metallicity}

Figure \ref{metal} shows the relationship between the gas 12 $+$ Log(O/H) 
index and the total H band luminosity of galaxies in the Virgo cluster.
Given the large uncertainty in both models 
(it is well known that metal yields are uncertain by at least a
factor 2, see e.g. Prantzos 2000, and chemical evolution models are somewhat degenerate due to
uncertainties on e.g. the Initial Mass Function or star formation history)
and observables (0.3 dex), it is not surprising that models for the unperturbed galaxies
are slightly shifted with respect to the data. We should thus
consider only relative values. It is however clear that
all starvation and ram pressure models predict both an increase of the
gas metallicity and a decrease of the total H band luminosity
consistent with the dispersion in the data.
This might explain the increase of gas metallicity observed in Virgo (Vilchez 1995)
and Hydra (Duc et al. 2001) cluster galaxies.

\subsection{Structural properties}

The structural properties of quiescent and star forming dwarf galaxies are 
compared to model predictions in Fig. \ref{c31} and \ref{binggeliLH}.
The near-IR light concentration index $C_{31}$ (defined as the
ratio of the radii including 75 and 25\% of the total light), is a quantitative
indicator of the shape of the light profile in galaxies 
which reflects the $n$ exponent of the Sersic law. In practice it can be used as an indicator of
the presence of bulges 
\footnote{The concentration index is however insensitive to the presence of 
small nuclei as those present in nucleated Virgo cluster dE},
thus to quantitatively compare the radial profiles
of star forming and quiescent dwarfs, as shown in Fig. \ref{c31}. 
For luminosities $L_H$ $\leq$ 10$^{8.4} $L$_{H \odot}$ the quiescent galaxy population
is undersampled. In the 10$^{8.4}$ $<$ $L_H$ $\leq$ 10$^{9}$ L$_{H \odot}$
both galaxy populations are dominated by objects with 
near-IR concentration indices $C_{31}(H)$ $\leq$ 3 (55\% and 78\% for early and late-type
respectively), typical of exponential profiles.
The frequency of dwarfs with pronounced central cusps
($C_{31}(H)$ $>$ 3) increases with luminosity, with an overall frequency comparable
in the quiescent (64\%) and star forming (42\%) galaxy populations in the luminosity range 
10$^{9}$ $<$ $L_H$ $\leq$ 10$^{9.6}$ L$_{H \odot}$. 
Values of $C_{31}(H)$ $>$ 3 are more frequent in nucleated dEs (large filled dots), 
and in high surface brightness star forming galaxies (spirals, black open squares and green crosses, 
and BCDs\footnote{We included in this category also the
mixed classes S/BCD, Sm/BCD and Im/BCD}, blue open triangles). 
We remind that ram pressure stripping events are not expected to deeply modify the relative weight of bulges in 
perturbed galaxies (Boselli et al. 2006).\\
For the star forming galaxies, the H band effective surface 
brightness and the luminosity are correlated, with a relatively large
scatter, with higher disk surface brightnesses in more massive galaxies. 
The unperturbed models predict a similar correlation, but with 
a small shift with respect to the observations (which show higher surface brightnesses 
than model predictions). These discrepancies
could be explained by a combination of several effects.
First, as we noted above, the models are computed for an average spin, but
surface brightness is sensitive to the spin parameter, thus we do expect a
large scatter in such a plot. Second, small galaxies may have intrinsic 
irregularities (linked to sporadic episodes of star formation) affecting
the effective surface brightness, which are not included in the models. Finally,
the models only consider pure disks, which is the case only at low luminosities. 
If we make a further separation of the star forming dwarfs into 
different morphological classes (see Fig. \ref{binggeliLH}) we can see that
the lowest surface brightnesses are reached only by magellanic irregulars (Sm-Im).
The H band effective surface brightnesses of the dE are similar to those of 
the star forming BCD and spirals (Sa-Sd) of similar luminosity. 
We notice however that the sample of dE with available near-IR imaging
is slightly biased towards high surface brightness, nucleated
objects: among the 15 dE galaxies with the lowest values of $\mu_e(B)$, only 20\% have
a measure of $\mu_e(H)$, while among B band high surface brightness objects
69\% have measured $\mu_e(H)$. \\
The relationship between the B band effective surface brightness and
the H band luminosity of dE and Scd-Im-BCD is significantly dispersed,
but consistent with the model predictions. 
The effective surface brightnesses of quiescent and star forming 
systems are more similar in the B band than in the H band. We notice however that dwarf ellipticals have 
effective surface brightnesses slightly higher than magellanic irregulars and 
smaller than late-type spirals and BCD of similar luminosity.
Quiescent systems resulting from the quenching of the star formation activity
of magellanic irregulars, whose existence is predicted by our models, are 
excluded from our sample just because they have B band surface brightnesses below the
detection limit of the VCC. 
\\
The variations induced by the interactions are more evident
in the UV bands. In the NUV band, for instance, star forming 
galaxies (of all types) have $\sim$ 
2-3 magnitudes brighter surface brightnesses than quiescent dwarfs, as predicted by models.
The most important variations of the surface brightness of perturbed galaxies
predicted by models are in the FUV band (up to 4 magnitudes). The low GALEX sensitivity, 
however, prevents us to detect
most of the dE that, as expected, have $\mu_e(FUV)$ $>$ 26.5 mag arcsec$^{-2}$.\\
\\
We can thus conclude that the observed spectrophotometrical and structural properties of dwarf ellipticals 
are consistent with those of star forming dwarfs once their activity has been stopped 
after a ram pressure gas stripping event able to remove their total gas content. 
High surface brightness star forming objects should be the
progenitors of the dwarf ellipticals, while magellanic irregulars the progenitors of
quiescent objects of low surface brightness not detected in the VCC.\\

\section{The transitional objects}

\begin{figure*}
\epsscale{1.0} \includegraphics[width=15cm,angle=0]{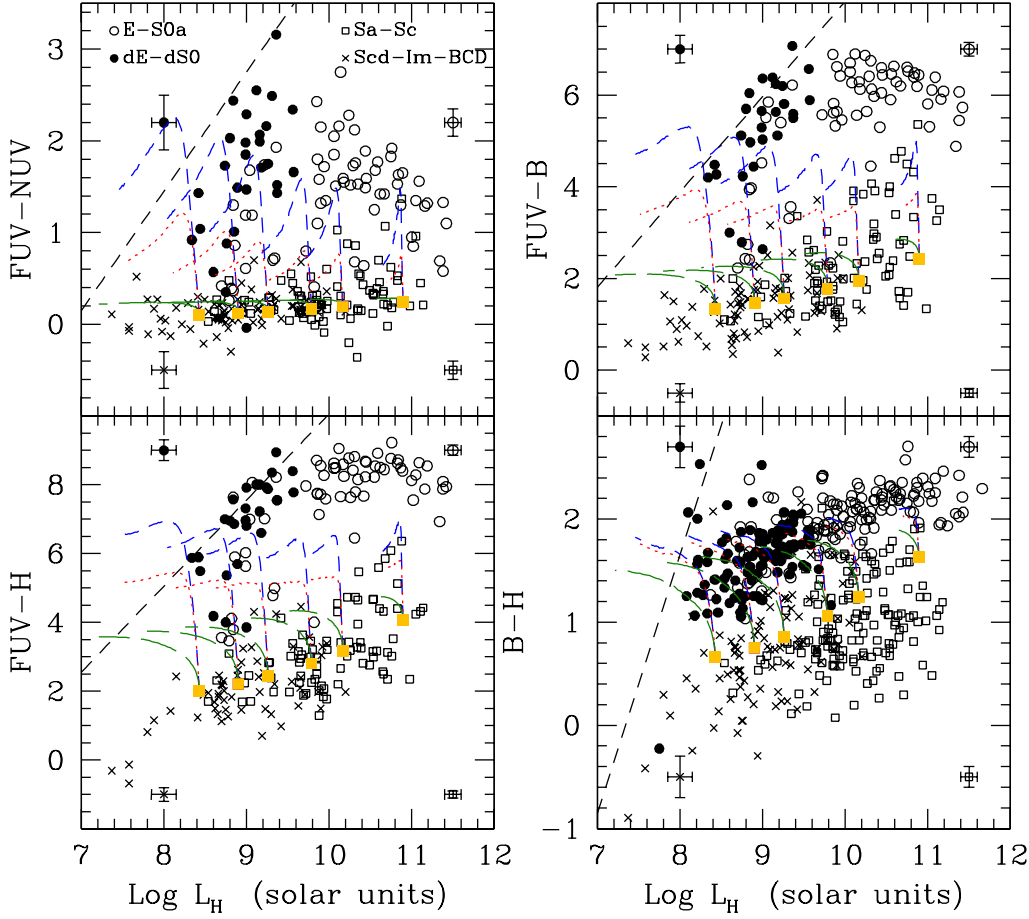}
\small{\caption{The relationships between different UV, optical and near-IR color indices 
and the H band luminosity compared to model predictions. 
Symbols and models are as in Fig. \ref{def}. Galaxies redder than the dashed line are undetectable
by the present surveys.
\label{colmag}}}
\end{figure*}

\begin{figure*}
\epsscale{1.3} 
\plotone{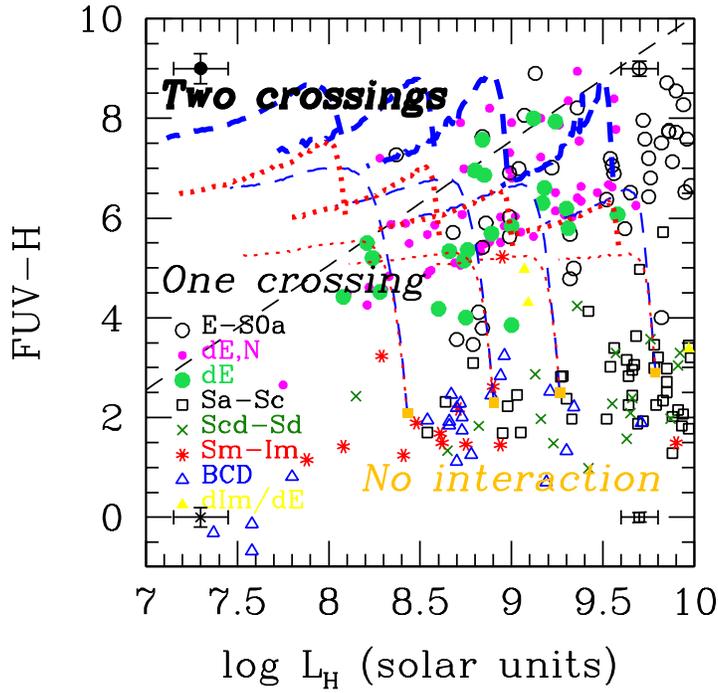}
\small{\caption{The relationships between the $FUV-H$ color index 
and the H band luminosity of dwarf galaxies compared to model predictions (in blue for an efficiency of 
$\epsilon_0$ = 0.4 M$\odot$ kpc$^{-2}$ yr$^{-1}$ and red $\epsilon_0$ = 1.2 M$\odot$ kpc$^{-2}$ yr$^{-1}$)
in the case of single (thin lines) or multiple (thick lines)
interactions. Symbols are: black empty squares for Sa-Sc, dark green crosses for Scd-Sd, red asterisks for Sm-Im,
blue open triangles for BCD, yellow filled triangles for dIm/dE, black open circles for E-S0a, magenta small filled dots
for nucleated dE and green large filled dots for non-nucleated dE.  
Models are for galaxies of spin parameter $\lambda$ = 0.05  and 
rotational velocity $V_C$ = 40, 55, 70 and 100 km s$^{-1}$ from left to right, respectively.
Galaxies redder than the dashed line are undetectable by the present surveys.
\label{colmag_2}}}
\end{figure*}

\begin{figure*}
\epsscale{1.3} 
\plotone{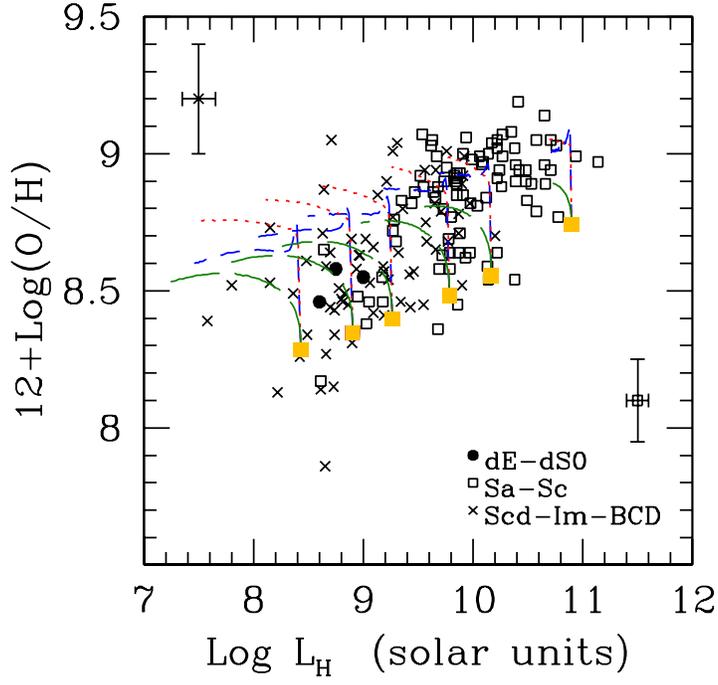}
\small{\caption{The relationships between the gas 12 + Log(O/H) 
metallicity index and the model predictions. 
Symbols and models are as in Fig. \ref{def}. The three dE galaxies 
in the plot are the transitional types VCC 710 (12 + Log(O/H) = 8.55), VCC 1501 (12 + Log(O/H) = 8.46)
and VCC 1617 (12 + Log(O/H) = 8.58). 
\label{metal}}}
\end{figure*}

\begin{figure*}
\epsscale{1.0} \includegraphics[width=15cm,angle=0]{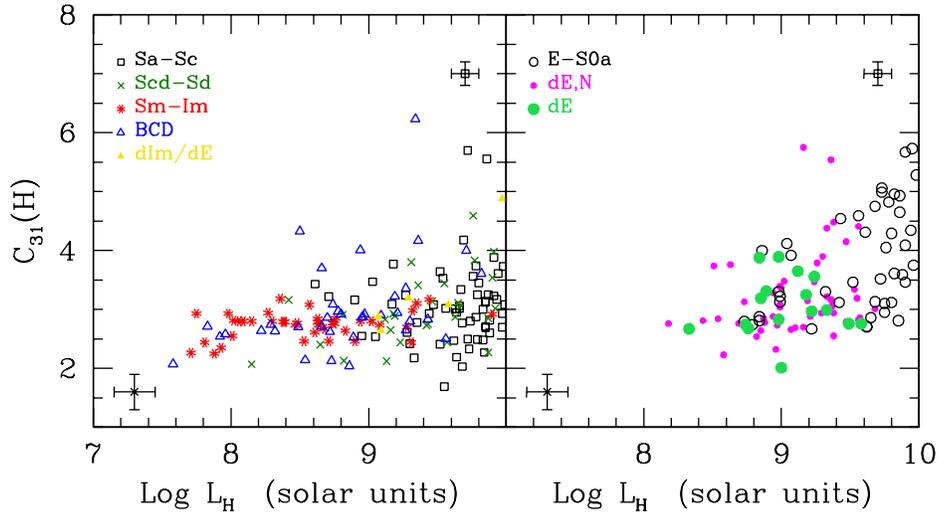}
\small{\caption{The relationship between the H band $C_{31}(H)$ parameter and the H band luminosity for early- (right) 
and late-type (left) galaxies 
with $L_H$ $<$ 10$^{10}$ L$_{H \odot}$. Symbols are as in Fig. \ref{colmag_2}.
\label{c31}}}
\end{figure*}

\begin{figure*}
\epsscale{1.0} \includegraphics[width=15cm,angle=0]{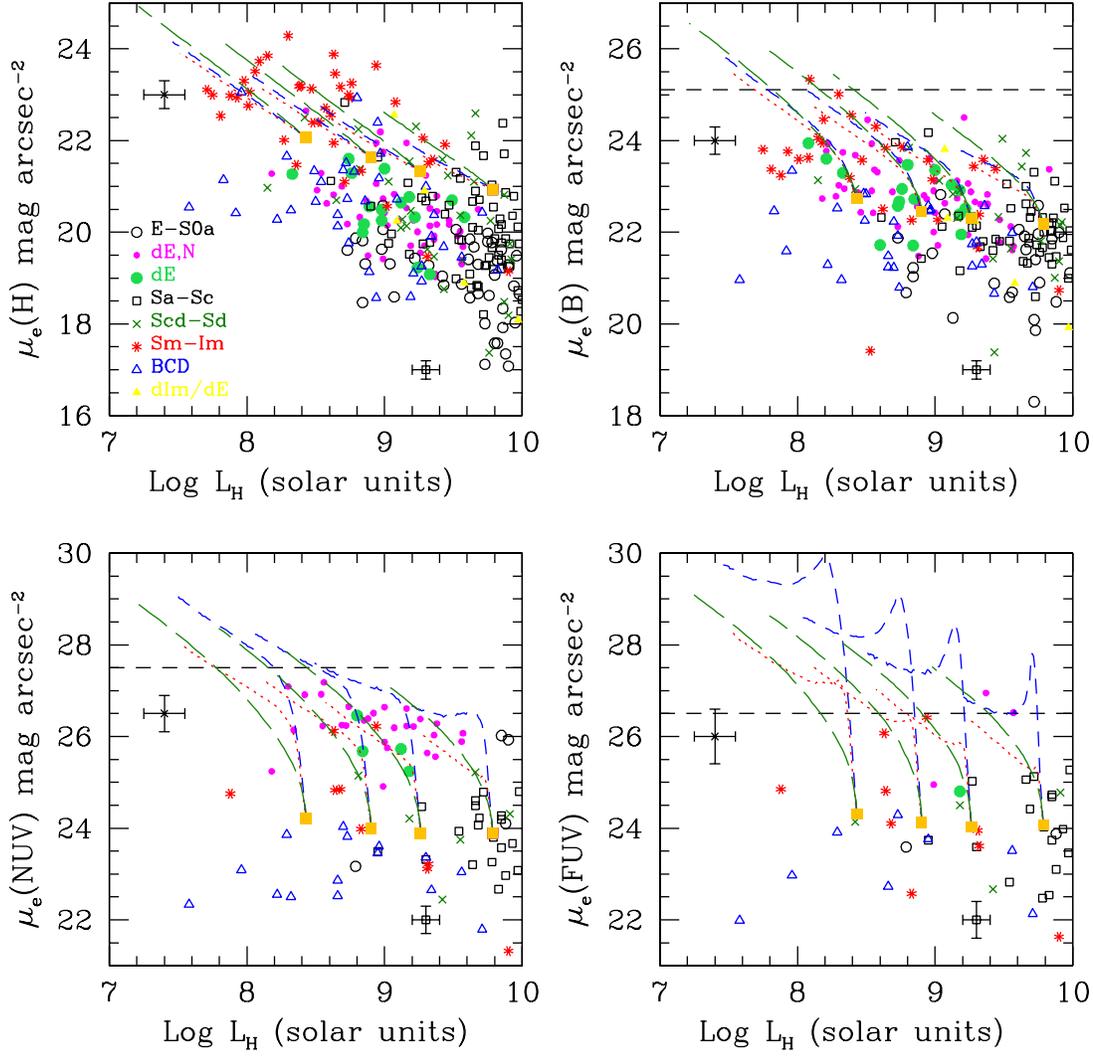}
\small{\caption{The relationships between the H, B, NUV and FUV effective surface brightnesses (in AB magnitudes)
and the H band luminosity. Symbols and models are as in Fig. \ref{colmag_2}. The horizontal dashed lines
in the B, NUV and FUV panels are the surface brightness detection limits of the VCC and GALEX.
\label{binggeliLH}}}
\end{figure*}

\begin{figure}
\epsscale{1.0} \includegraphics[width=15cm,angle=0]{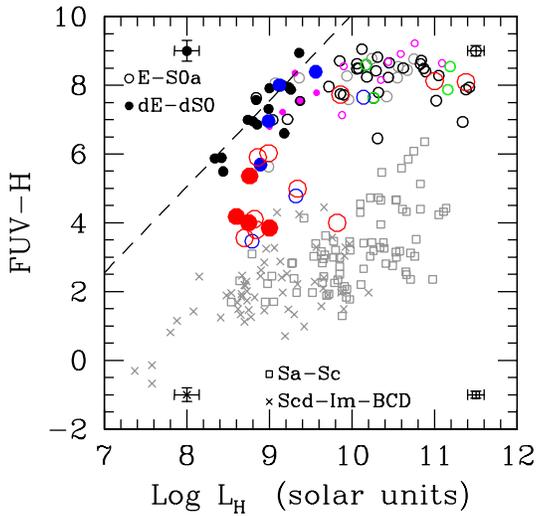}
\small{\caption{The relationship between the FUV-H color index 
and the H band luminosity (see Fig. \ref{colmag}).
Early-type galaxies (filled circles for dE-dS0, empty circles for E-S0-S0a) are coded as follow:
red symbols are for galaxies with H$\alpha$ emission (H$\alpha$E.W.$_{em}$ $>$ 2 \AA), blue symbols 
for objects with strong H$\beta$ absorption lines (H$\beta$E.W.$_{abs}$ $>$ 2.8 \AA), green symbols 
for HI detected galaxies, magenta symbols for 
rotationally supported galaxies ($v^*/\sigma) > 0.5$), black symbols for galaxies with spectroscopy 
but not satisfying the previous criteria. Colors are mutually exclusives,
with priority ordered according first to the H$\alpha$ emission, then to the H$\beta$ absorption, to
the HI detection and to rotationally supported objects. Gray symbols are for the whole late-type galaxy population 
and for early-type objects without any spectroscopic information.
\label{colmag2}}}
\end{figure}

\begin{figure}
\epsscale{1.0} 
\includegraphics[width=8cm,angle=0]{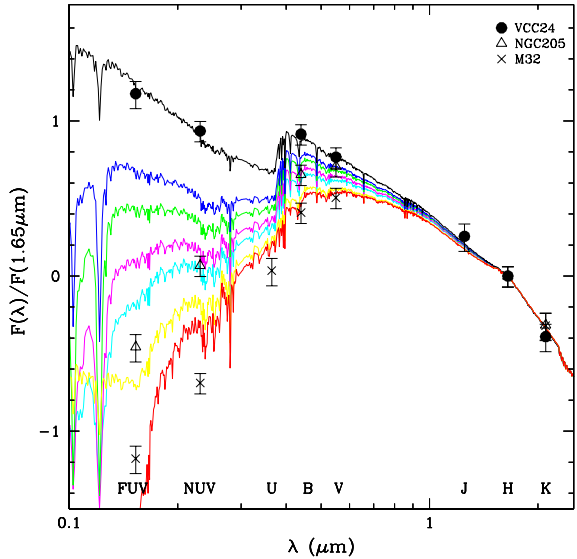}
\small{\caption{
The UV to near-IR spectral energy distributions of the BCD VCC 24 (filled dots), of the dE NGC 205 (empty triangles) 
and the compact dwarf elliptical M32 (crosses) (all normalized to the 1.65 $\mu$m H band flux) are compared to 
the model SED for a galaxy of spin parameter $\lambda$ = 0.05
and rotational velocity $V_C$ = 55 km s$^{-1}$. The black continuum line is the model for an unperturbed galaxy, the continuum blue, 
green, magenta, cyan, yellow and red are for galaxies which had their peak of the interaction 
respectively 0, 100, 300, 500, 1300 Myr and two cluster crossings with the last one 
500 Myr ago.
\label{sed_ref}}}
\end{figure}

The analysis of Fig. \ref{colmag} (see previous section) reveals the existence of several objects having 
spectro-photometric properties in between that of star forming late-type galaxies 
and quiescent spheroidals.
The UV selection, which is strongly biased towards star forming objects,
favors the detection of such blue, still active dwarf ellipticals.
Figure \ref{colmag2} shows the same FUV-H color-magnitude relation given in Fig. \ref{colmag} 
but using different colour symbols (only for early-types)
to indicate still star forming galaxies (H$\alpha$E.W.$_{em}$$>$ 2 \AA, red), post-starburst galaxies which
recently ended their activity as indicated by their strong Balmer H$\beta$ absorption line 
(H$\beta$E.W.$_{abs}$$>$ 2.8 \AA, blue), HI detected objects 
(green) and rotationally supported spheroidals ($v^*$/$\sigma$$>$0.5, magenta)
\footnote{The anisotropy parameter $v^*$/$\sigma$ is defined as: 
$v^*$/$\sigma$=$(v/\sigma)/[e/(1-e)]^{1/2}$, where $e$ is the ellipticity of the galaxy} .

We can photometrically define the transitional class as being formed by those objects 
of luminosity Log $L_H$ $<$ 9.6 L$_{H \odot}$ with a 3$<$ $FUV-H$ $<$ 6. 
The early type galaxies in our sample satisfying this criteria are listed in Table \ref{Tabtrans}.

\subsection{The UV to near-IR spectral energy distribution}

\begin{figure}
\epsscale{1.0} 
\includegraphics[width=15cm,angle=0]{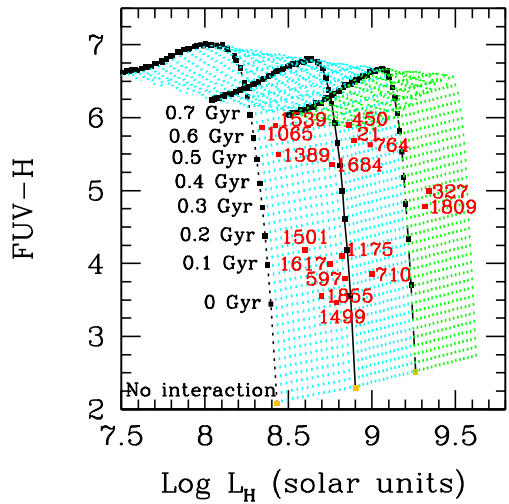}
\small{\caption{
The transition objects are shown in the FUV-H versus H band luminosity plane.
The curves indicate the position of various ram pressure stripping models 
($\epsilon_0$ = 1.2 M$\odot$ kpc$^{-2}$ yr$^{-1}$) interpolated between
40 and 70 km s$^{-1}$ (and extrapolated beyond) without interaction (orange squares), and
for various lookback time to the interaction (a few are indicated on the left).
These models were used to fit a velocity and lookback time for each
transition objects.
\label{sed_fit}}}
\end{figure}

Figure \ref{colmag2} and Table \ref{Tabtrans} clearly show that most of the low luminosity early-type galaxies with blue colors are 
characterized by an ongoing or recent star formation activity (strong emission or absorption Balmer lines; see Fig. \ref{sed}). 
Two of them have been detected in HI, while none have rotational velocity measurements. \\
Figure \ref{sed_ref} shows how the UV to near-IR spectral energy distribution of a star forming dwarf galaxy of $\lambda$ = 0.05 and 
rotational velocity $V_C$ =  55 km s$^{-1}$ changes with time after a ram pressure stripping event characterized by an efficiency 
$\epsilon_0$ = 1.2 M$\odot$ kpc$^{-2}$ yr$^{-1}$. For comparison the model SED are plotted along with 
the observed SED of the BCD galaxy VCC 24 and the dwarf ellipticals M32 and 
NGC 205\footnote{We compare the prediction of our 
model to the SED of NGC 205 and M32 just because these two objects are the 
only well known dwarf ellipticals detected by GALEX in the two filters. 
We remind however that NGC 205 is a non standard dE with
central gas and some star formation activity that would make it classified as
a dE with blue center by Lisker et al. (2006b). Given its high surface brightness,
M32 is a non genuine dwarf elliptical. The photometric data of NGC 205 and M32 are taken from Gil de Paz et al. (2007)}.
The SED of an unperturbed galaxy is similar to that 
of star forming dwarf systems such as the BCD VCC 24, while it becomes redder soon after the interaction, reaching the values of 
the dwarf elliptical NGC 205 after 1.3 Gyr. Two interactions are necessary to get colors as red as those of 
the compact dwarf elliptical M32.\\
%
During the transition phase, the FUV-H colour gets redder while the
H band magnitude is slightly reduced with respect to the reference
case (Fig. \ref{sed_fit}). We can use these properties to estimate the
rotational velocity and lookback time to the interaction. All the galaxies are
close to the 40-55-70 km s$^{-1}$ range for which we computed
models. We thus interpolated (and extrapolated above 70 km s$^{-1}$) in
1 km s$^{-1}$ bins our models. Fig. \ref{sed_fit} shows the position of
each of these models (for the 1.2 M$\odot$ kpc$^{-2}$ efficiency) in
the $FUV-H$ vs Log $L_H$ diagram. 
The models at the bottom of the
diagram are unperturbed. Each point along the lines for different
velocities correspond to various value of the stripping look-back time
($t_{rp}$=0, 0.1, 0.2, 0.3 ...).  Velocities and
look back times for each transitional object can easily be found by a
simple least-square test (see Table
\ref{Tabtrans} and Fig. \ref{sed}). Models indicate that all transitional objects recently
($<$ 650 Myr) interacted  with the cluster IGM.

\subsection{Color gradients}

It has been shown that some dwarf elliptical galaxies have inverted
color gradients, with bluer colors in the center (Lisker et al.  2006b
and references therein). Indeed six out of the sixteen galaxies listed
in Table \ref{Tabtrans} have been defined as ``blue center'' by Lisker
et al. (2006b). \\
The observed inversion of the color gradient in the center of dwarf
galaxies is obtained in our multizone models in a ram pressure
stripping scenario (see Fig. \ref{profiles}) for recent interaction of relatively
weak efficiency (0.1 M$\odot$ kpc$^{-2}$), and has the same origin
than that observed in massive spirals such as NGC 4569 (Boselli et
al. 2006). Blue colors in the center of these dwarf galaxies are
expected because ram pressure stripping is more efficient in the outer
disk, where all the gas is removed, while some gas needed to feed star
formation might be retained in the central part where the galaxy
potential is deeper. 
For more efficient interactions, the contribution of the recycled
gas becomes important whenever the interaction started long ago 
in the center of galaxies, where the stellar surface
density is high and we obtain again blue centers. 
We should note that if galaxies have deep potential wells because 
of a central mass condensation (not taken
into account in our models) the gas in the center should be harder to remove, 
easily making blue cores.
Given the shape of the potential well, whose deepness
increases with mass (Catinella et al. 2006), we expect that the
inversion of the color gradient is more frequent and radially extended
in the brightest dwarfs, which are more likely to retain some gas,
as indeed observed (Lisker et al. 2006b).
The lookback time to the ram pressure stripping event, combined with
the age of the different stellar populations sampled in the various
filters, make the observed colour inversion more pronounced at short
wavelengths, as shown in Figure \ref{profiles}. We can also add that
such an inversion of the colour gradient can not be reproduced in a
starvation scenario where gas removal is not a radial but rather a
global effect.\\
For those galaxies with available SDSS images we reconstructed their $NUV-i$ colour gradient (see Fig. \ref{sed}) using
SDSS images smoothed to the GALEX resolution. Twelve out of the 14 galaxies with available data have indeed 
blue colors in their center (Fig. \ref{sed} and Table \ref{Tabtrans}), as expected if these galaxies 
recently underwent a ram pressure stripping event. We remark that the observed blueing of the inner gradient happens 
in the inner 1-2 kpc, as predicted by models (Fig. \ref{profiles}).\\
Seven of the transitional class galaxies (VCC 327, 450, 597, 710, 1175, 1617 and 1855) 
have been imaged in H$\alpha$ with the San Pedro Martir 2m telescope (Gavazzi et al. 2006a).
All galaxies (except VCC 327, undetected in H$\alpha$ imaging) show a similar pattern, with
the star formation activity always limited to the central region and 
not extended to the whole disk, once again as predicted by our model, as shown in Fig. \ref{ha} for VCC 710. 

\section{Discussion}

The analysis done so far clearly indicates that some dwarf ellipticals might result from the rapid decrease of the star
formation activity of low luminosity, late type galaxies after most of their gas was lost because of their strong 
interaction with the hostile cluster environment. The comparison between model predictions and observations favors ram
pressure stripping events with respect to galaxy starvation. Models in fact predict 
that low luminosity spirals, Im and BCD suffering starvation since 6 Gyr are still star forming galaxies 
with blue UV colors, relatively strong Balmer emission lines (H$\alpha$E.W.$_{em}$ $\sim$ 10 \AA) and have an 
HI gas content ($\sim$  10$^8$  M$\odot$) easily detectable by HI surveys such as ALFALFA and AGES.
These objects would thus still be classified as star forming dwarf galaxies, and not dwarf ellipticals.\\
We should first notice that our simple representation of a ram pressure stripping event is totally consistent with the  
high resolution hydrodynamical simulations in cluster dwarf galaxies of Mori \& Burkert (2000) 
Murakami \& Babul (1999) and Marcolini et al. (2003), which predict the stripping of the whole gas reservoir on short timescales.\\ 

\subsection{Is the ram pressure scenario consistent with all observational evidences?}

As mentioned in the introduction, however, several observational evidences, namely 
1) the different structural properties of magellanic irregulars and dwarf ellipticals, 2) the presence of nuclei
in dE, 3) the higher specific frequency of globular clusters and 4) the higher stellar metallicity 
of quiescent systems with respect to star forming dwarfs, 5) their different velocity distribution and 6)
the different shape of the field and cluster luminosity functions have been often indicated as
major limitations to the formation of cluster dwarf ellipticals through the gas stripping of star forming systems.
We see in this section how these difficulties can be overcome.

\subsubsection{Structural properties}

In general we can say that the present work gives a natural explication to the presence 
of rotationally supported (Pedraz et al. 2002; Geha et al. 2003; van Zee et al. 2004b), gas rich 
(Conselice et al. 2003b; van Zee et al. 2004b) star forming dwarf ellipticals. 
The presence in some dwarf ellipticals of spiral arms observed by Lisker et al. (2006a) is a further evidence of the
rotating/spiral origin of these objects. The grand design spiral arms
of these dE, apparently different from the flocculent open arms of very late-type spirals, might be driven by  
disk instabilities produced by the displacement of the gas over the stellar disk during the 
stripping event (Elmegreen et al. 2002). Consistently with observations,
which have shown that the frequency of dE with grand design spiral arms increases with luminosity (Lisker et al. 2006a), 
the perturbations induced by the gas displacement are expected to be important only in 
those objects where the potential is sufficient to retain some gas, while totally absent in the 
lowest luminosity systems where the whole gas content is instantaneously removed. \\
The comparison of model predictions with observations suggests 
that on average the variation of the structural parameters (surface brightnesses, effective radii 
and concentration indices) of high surface brightness low-luminosity star forming 
galaxies such as BCD or late-type spirals (mostly Scd-Sd)
at various wavelengths expected after a ram pressure stripping event might indeed 
reach the values observed in dE. Consistently with Bothun et al. (1986)
we confirm that the sampled dE have near-IR and optical surface brightness higher than those
of magellanic irregulars (Sm-Im) and thus cannot result from this galaxy population.
Our models, however, predict the existence in the Virgo cluster of an extremely 
low surface brightness quiescent galaxy population ($\mu_e(B)$ $\geq$ 25 mag arcsec$^{-2}$) 
resulting from the quenching of the star formation activity of magellanic irregulars. Such a population has been
indeed observed by Sabatini et al. (2005). 
%
\\
Our model predictions can not be easily compared to the results of Lisker et al. (2006a,b; 2007) and Binggeli \& Popescu (1995)  
on the flattening distribution of star forming and quiescent dwarfs since models do not take into account the 3-D evolution of the disks.
We can argue that the lack of supply of young stars with low velocity dispersion because of the suppression of the
star formation activity causes the disk to heat up, dumping spiral waves on timescales of a few revolutions (Sellwood \& Calberg 1984; Fuchs \& von Linden
1998; Elmegreen et al. 2002) in particular in low luminosity systems where the rotational component is not always dominant.
In low luminosity rotating systems (70-130 km s$^{-1}$) it has been observed that the disk scale height increases by $\sim$ 
a factor of 2 in $\sim$ 3 Gyr (Seth et al. 2005). We could thus expect that the flattening distribution is
age dependent. In the lack of external gravitational perturbations, the rotational velocity is conserved while 
$\sigma$ increases once the gas is removed. Gravitational perturbations induced by the interaction with other galaxies 
or with the cluster potential would heat up the system and disperse the angular momentum, thus speeding up 
the decrease of $v/\sigma$ (Wozniak 2007, private communication).
We would thus expect rounder shapes and more relaxed distributions in those objects where the star formation stopped 
long time ago. We remark that, although still not virialized, nucleated objects are more 
centrally clustered than non nucleated systems (Lisker et al. 2007). 



\subsubsection{Nuclearity}

The angular resolution of our model is too poor ($\sim$ 1 kpc)
to make any prediction on the formation process and evolution of the nuclei
observed in dwarf ellipticals. As a general remark we can say that
recent ACS/HST high resolution observations of nuclei of early-type galaxies in the Virgo cluster
revealed that they have statistical (frequencies) and physical (sizes, luminosities, colors) 
properties similar to those of the nuclear stellar clusters found in late-type galaxies (C\^ote et al. 2006),
consistently with our evolutionary picture.
HST observations of dE
show color gradients getting bluer towards their nuclei witnessing a younger
stellar population than that of the outer disk (Lotz et al. 2004; C\^ote et al. 2006).
Furthermore the presence of a nucleus is more frequent in
massive dE than in low luminosity objects (Sandage et al. 1985;
Ferguson \& Binggeli 1994)\footnote{High resolution HST 
observations of dwarf ellipticals in the Virgo cluster 
revealed the presence of small nuclei also in objects classified as non-nucleated in the VCC 
(Lotz et al. 2004; C\^ote et al. 2006), whose distribution
does not seem to depend on the position within the Virgo cluster. For consistency with our morphological
classification and for completeness reasons,
we consider here the VCC classification which might be biased, for 
its limited angular resolution, to the most extended nuclei.}. 
\\

\subsubsection{Specific frequency of globular clusters}

\noindent
The rapid evolution expected after the galaxy-ICM interaction which abruptly stops the star formation is able to change the
total luminosity of the perturbed galaxy. The typical progenitor of a given dwarf elliptical 
that did not undergo such a violent truncation of its star forming activity today would be much brighter than its potential 
progeny even in the near-IR (see for instance Fig. \ref{colmag}). 
If we make the reasonable assumption that globular clusters 
were formed in the early phase of galaxy formation (thus well before any possible recent ram pressure stripping event) 
and regulated by the primordial dark matter condensation, the specific frequency of globular cluster should thus be determined by
normalizing the total number of globular clusters to the absolute magnitude of the unperturbed galaxy which is always brighter than its 
gas stripped counterpart. Being the specific frequency of globular clusters defined as 
$S_N$ $=$ $N_{GC}$ $\times$ 10$^{0.4(M_V+15)}$ (where $N_{GC}$ is the number of globular cluster; 
Strader et al. 2006), we can see that $S_N$ increases with time after a ram pressure stripping 
event just because the absolute magnitude $M_V$ of the perturbed system does not increase with time as
that of its unperturbed counterpart (see Fig. \ref{GC}).
The specific frequency of globular clusters in dE galaxies ranges in between $\sim$ 1 and $\sim$ 20 with $\sim$ a double 
distribution peaked at $S_N$ $\sim$ 2 and $S_N$ $\sim$ 10 (Strader et al. 2006).
These values are consistent with a scenario where dE have been formed through a ram pressure gas stripping event
of low luminosity late type galaxies (whose specific frequency is $S_N$ $\leq$ 1; Miller et al. 1998)
occurred $\leq$ 4 Gyr ($\leq$ 1 Gyr for dE with $S_N$ $\sim$ 2).
\\

\subsubsection{Metallicity}

Grebel et al. (2003) observed an increase of the 
metallicity measured in the old stellar populations in dwarf
spheroidals with respect to dwarf irregulars of similar luminosity.
It is equivalent to say that the spheroidals are fainter than
irregulars for the same stellar metallicity.
This is naturally obtained in our models: let's consider a ``common
ancestor'', giving rise to a low luminosity, star forming galaxy if it evolves in
isolation, and to a dE if its star formation activity is suppressed
after an interaction.
The metallicity in the old population will be the same (the one of
the common ancestor). The luminosity of the star forming object will be
the one of the unperturbed model, while the one of the quiescent system
will be the one of the model after an interaction. Fig.
\ref{GC} shows that the difference in V band luminosity (partly due to the
fading of the stellar population, but also to the fact that less
stars formed in the latter than in the former case) is of $\sim$ 1
magnitude after 1 Gyr. This luminosity difference for the same
metallicity corresponds well to that observed by Grebel et al. (2003) between
the star forming and quiescent dwarf galaxy populations.
\\

\subsubsection{Velocity distribution}

The velocity distribution of the low luminosity ($L_H$ $<$ 10$^{10}$ L$_{H\odot}$), high surface brightness ($\mu_e(H)$$<$ 22 mag arcsec$^{-2}$) 
star forming galaxies (see Table \ref{Tabvel}), i.e. of those objects that our models indicate as the probable progenitors of the observed dwarf ellipticals, 
is strongly non gaussian, with a pronounced wing 
at high velocity (see Fig. \ref{vel}), witnessing infall. If limited to the Virgo A subcluster, thus to the region associated to the
hot gas emitting in X-ray where ram pressure is active (Gavazzi et al. 1999), the probability that both nucleated and non nucleated
dwarf ellipticals are driven by the same parent population as low luminosity, high surface brightness star forming galaxies
is relatively important ($P$ $\sim$ 40\%), as indicated by Kolmogorov-Smirnov tests (see Table \ref{TabKS}).
Although dynamical considerations are consistent with the infall of the whole dwarf elliptical galaxy population, 
it is however difficult to determine when this happened: the velocity dispersion inside the cluster
of their progenitor star forming, low-luminosity galaxy population should not change very much over a Hubble time as a result of dynamical friction, 
relaxation or energy equipartition as shown by Conselice et al. (2001).
\\

\subsubsection{Luminosity function}

The scenario where all Virgo cluster early type dwarfs have been created after the gas removal of low-luminosity late type
galaxies recently infalled into the cluster is however too extreme since quiescent dwarf galaxies are present also in the field
(Pasquali et al. 2005).
The contribution to the faint end of the field luminosity function of dE, however, is less important than that of blue, star forming
objects (Blanton et al. 2005), it is thus possible that the fraction of dwarf galaxies recently perturbed by the cluster environment
is very high. We also remark that the slope of the Virgo cluster luminosity function ($\alpha$ $\sim$ -1.4; Sandage et al. 1985)
is comparable to the most recent determinations of the field luminosity function based on SDSS data,
$\alpha$ $\sim$ -1.4, -1.5 (Blanton et al. 2005). Steeper slopes in other clusters luminosity functions have been however determined
by Popesso et al. (2006).\\

\subsection{Infall rate}

The question now is to see how many Virgo cluster dwarf ellipticals might have been formed after ram pressure stripping of
low-luminosity star forming systems. 
To answer this fundamental question we should know both the infalling rate of low luminosity, star forming galaxies
in cluster and the time scale for a galaxy to totally stop its activity. 
Our models give us an idea of the different time scales for gas removal and suppression of the
star formation and of the subsequent evolution of the different stellar populations inhabiting these galaxies. 
Figure \ref{clock} shows how, on relatively short time scales, the HI deficiency parameter, 
on two different indicators tracing the current (H$\alpha$ emission line; 
Kennicutt 1998; Boselli et al. 2001) and past (H$\beta$ absorption line\footnote{This is the  
absorption line uncontaminated by the H$\beta$ emission, which might be present 
whenever star formation is still active}; 
Poggianti \& Barbaro 1997; Poggianti et al. 2001a, 2001b; Thomas et al. 2004)
star formation activity and the
$FUV-H$ color index are expected to evolve after a ram pressure stripping event for galaxies with $L_H$ $<$ 10$^{9.6}$ L$_{H\odot}$.
\noindent
Gas removal is extremely efficient in such low-mass objects, making the HI mass decrease by $\sim$ two orders of magnitude
on very short time scales, i.e. $\sim$ 150 Myr (see Table \ref{Tabstat}). The lack of gas causes, on similar time scales, a rapid
decrease of the H$\alpha$E.W.$_{em}$ (we remind that the H$\alpha$ emission of a galaxy is due to the gas ionized by O-B stars of ages $<$
10$^7$ years, Kennicutt 1998). The interaction is so efficient that both indicators are already strongly perturbed even before the
galaxy reaches the center of the cluster (lookback time to the interaction=0 in Fig. \ref{clock}). The H$\beta$E.W.$_{abs}$ and the $FUV-H$ color
index, on the other hand, evolve more gradually in time: we can say that on average the equivalent width of the H$\beta$ 
absorption line drops down to $<$ 2.8 \AA ~ after $\sim$ 0.5-0.8 Gyr, while $FUV-H$ becomes redder by 5 mag after $\sim$ 0.8 Gyr
independently from the adopted ram pressure stripping model.
The transition in between the blue sequence of low-luminosity star forming galaxies 
and the red sequence of quiescent dwarf ellipticals is thus a
very rapid event, it is thus not surprising that only a few objects populate the intermediate region. \\

Table \ref{Tabstat} indicates that the fraction of low-luminosity early type galaxies in the absolute
magnitude range -13.15$>$ $M_B$ $>$ -17.5 
(this upper limit roughly corresponds to $L_H$ $<$ 10$^{9.6}$ L$_{H \odot}$) 
with a residual star formation activity (H$\alpha$E.W.$_{em}$ $>$ 2 \AA), still rich in HI gas, with a post 
starburst activity (H$\beta$E.W.$_{abs}$$>$ 2.8 \AA) or with blue colors ($FUV-H$ $<$ 5) is important. These
values indicate that $\sim$ 10-16 \% of the quiescent dwarfs might result from star forming systems which
underwent an interaction with the cluster medium 
less than 150 Myr ago, while up to $\sim$ 40 \% in the last $\sim$ 500-800 Myr \footnote{SDSS, Michielsen et al. (2007)
and Geha et al. (2003) spectroscopic H$\alpha$ and H$\beta$ data 
are limited to the central 3-4 arcsec of the observed galaxies, and are thus not directly comparable to 
the model predictions which are relative to the whole galaxy. 
Integrated spectroscopy however is available only for a small fraction of the quiescent
dwarfs.}. These values can be used to infer a dwarf galaxy infall rate of $\sim$ 300 objects per Gyr
(see Tab. \ref{Tabstat}), a relatively important rate if we consider the total number of
dwarf members of Virgo ($\sim$ 650, see sect. 2). This rate, however, is consistent with that
determined by Adami et al. (2005) for Coma scaled to the mass of Virgo: 
1-4 $\times$ 10$^{12}$ L$\odot$ Gyr$^{-1}$ if a mass to light ratio of 10 is adopted. 
\\

\subsection{Comparison with clusters at high z}

These values consistently indicate that the whole Virgo cluster dwarf quiescent galaxy population
might have been formed by the suppression of star formation in low luminosity, gas rich systems  
recently entered into the cluster if an infall rate similar to the present one lasted $\sim$ 2 Gyr 
(thus equivalent to $z$=0.16 in a $H_o$ = 70 km s$^{-1}$ Mpc$^{-1}$, $\Omega_M$=0.3 and $\Omega_{\lambda}$=0.7 cosmology).
This result is consistent with the most recent analysis of the color magnitude relation of clusters at different redshift
which all indicate a significant decrease of the fraction of low luminosity objects on the red sequence 
with increasing $z$ (De Lucia et al. 2004; 2007) from $z$=0.24 (Smail et al. 1998). It has also been shown
that the fraction of galaxies in between the red and the blue sequence of the color magnitude relation 
(which might be considered as the analogues of our transient population) in high density environment at $z$=0.7 from COSMOS 
increases with the decrease of the galaxy luminosity (Cassata et al. 2007), as predicted by our model.
Our results are also consistent with the analysis of Nelan et al. (2005) on 93 nearby clusters based on the age-$\sigma$
relation which indicates that the quiescent dwarf galaxy population only recently joined the red sequence.
\\

\section{Conclusion}

We can conclude that models and observations are consistent with an evolution of star forming, low-luminosity late type
galaxies recently accreted in Virgo into quiescent dwarfs because of the ram pressure gas stripping and the subsequent stopping of their star
formation activity. For consistency with surface brightness measurements, we show that high surface brightness
star forming dwarf galaxies (low luminosity spirals and BCD) might be at the origin of the optically selected dwarf ellipticals (both normal and nucleated) 
analyzed in this work, although we expect that the low surface brightness magellanic irregulars (Sm and Im)
once stripped of their gas reservoir, produce quiescent dwarfs with surface brightnesses below the detection limit of the VCC, as
those observed in Virgo by Sabatini et al. (2005). The process of transformation 
is extremely rapid and efficient, since it works on all dwarfs and last on average less than 150 Myr. On longer time scales galaxies get
structural and spectrophotometric properties similar to that of dwarf ellipticals. 
The whole star forming dwarf galaxy population dominating the faint end of the field luminosity function (Blanton et al. 2005),
if accreted, can be totally transformed by the cluster environment into dwarf ellipticals on time scales as short as 2 Gyr 
and thus be at the origin of the morphology segregation observed also at low luminosities.\\
This interesting result is of fundamental importance even in a
cosmological context because it shows that the majority of dwarf galaxies (if
not all of them) are ``young'' even in clusters, and not old as
expected in a hierarchical galaxy formation scenario (see also Nelan et al. 2005).  
For comparison with De Lucia et al (2006), we give in Table \ref{Tabage} the lookback time 
when 50\% and 80\% of the stars were formed for a model galaxy of $V_C$ $=$ 55 km s$^{-1}$ 
for a (or two) ram pressure stripping event of efficiency $\epsilon_0$ = 1.2 M$\odot$ kpc$^{-2}$ yr$^{-1}$.

\noindent
Although the values given in Table \ref{Tabage} are not directly comparable to those 
given by De Lucia et al. (2006) since limited to relatively high mass objects (2.5 10$^9$ M$\odot$,
while our model galaxy is of only 2.4 10$^8$ M$\odot$), the mean age of the stellar population 
of low luminosity, quiescent galaxies is significantly younger than that predicted by 
hierarchical models of galaxy formation (lookback times $\sim$ 10 and 8.5 Gyr
for 50\% and 80\% of the stellar population).\\
Despite their morphological type, the star formation activity of dwarf systems has been abruptly interrupted
by the interaction with the environment in quiescent systems. This
conclusion can be extended to the local group, where the study of the
stellar color magnitude relation of dwarf spheroidal systems revealed that the
star formation activity, although in an episodic manner, lasted for several Gyr (Mateo 1998; Grebel 1999). 
We can add that the only cluster galaxy population likely to be
issued by major merging events, as those predicted by hierarchical
models of galaxy formation, is that of massive ellipticals, whose
origin is probably very remote ($z$ $\geq$ 2-3, Dressler 2004; Treu
2004; Nolan 2004; Franx 2004; Renzini 2006). Indeed this is the only
Virgo cluster galaxy population with a virialized velocity distribution
(Conselice et al. 2001). If we consider clusters of galaxies as those
regions where merging events were more frequent at early epochs
since the big bang just because characterized by an high galaxy
density, we can conclude that the hierarchical formation scenario
was the principal driver of galaxy evolution only in massive objects at very early
epochs. All observational evidences are consistent with a
secular evolution afterward.

\acknowledgements 
We wish to thank C. Adami, E. Athanassoula, C. Balkowski, V. Buat, G. Comte, P.A. Duc, G. Hensler, J. Lequeux,
H. Wozniak for precious comments and suggestions. We are grateful to the anonymous referee 
for his constructive suggestions extremely useful
in the preparation of the final version of the manuscript.
GALEX (Galaxy Evolution Explorer) is a NASA Small
Explorer, launched in April 2003. We gratefully acknowledge NASA's
support for construction, operation, and science analysis for the
GALEX mission, developed in cooperation with the Centre National
d'Etudes Spatiales of France and the Korean Ministry of Science and
Technology. We wish to thank the GALEX
SODA team for their help in the data reduction.
This research has made use of the GOLD Mine database.


\begin{table*}
\caption{Effects of the interactions with respect to an unperturbed model.}
\label{Tabmod}
{\scriptsize
\[
\begin{tabular}{ccccccc}
\hline
\noalign{\smallskip}
Variable  	& \multicolumn{3}{c}{-------Ram~pressure-------}	& \multicolumn{3}{c}{-------Starvation-------}	    	\\
	  	& modification		& Amount	& Time~scale   	& modification		& Amount	 	& Time~scale$^*$\\
\noalign{\smallskip}	
\hline
\noalign{\smallskip}
Gas~content	&$\sim$~totally~removed	& factor~10-100	&$\la$150~Myr	& partly~removed	&$\la$~ factor~10	& 6~Gyr		\\
HI-deficiency	& strongly~deficient	& 1-2		&$\la$150~Myr	& mildly~deficient	&$\leq$0.5		& 6~Gyr		\\	
\hline
Colors		& strongly~reddened	&		&		& mildly~reddened	&			&		\\
FUV-NUV		& increased	  	& 1-2~mag	&$\la$1~Gyr	& increased		& 0.2~mag		& 6~Gyr	   	\\
FUV-H		& increased		& 3-5~mag	&$\la$1~Gyr	& increased		& 1.5~mag	    	& 6~Gyr	   	\\
FUV-B		& increased		& 2.5-4~mag	&$\la$1~Gyr	& increased		&$\la$1~mag  	    	& 6~Gyr	   	\\
B-H		& increased		& 1~mag	  	&$\la$1~Gyr	& increased		& 1~mag		    	& 6~Gyr	   	\\
color~gradients	& blue~center		& 1-2~mag~(NUV-i)& 1-5~Gyr	& red~center		& 1~mag			& any		\\
\hline
Star~formation	& stopped		&		&		& mildly~decreased	&			&		\\
H$\alpha$ E.W.	& drop~to		&$\sim$~0~\AA	&$\la$150~Myr	& drop~to		&$\sim$~10~\AA		& 6~Gyr		\\
H$\beta$ E.W.	& drop~to		&$\sim$~2.3~\AA	&$\la$1~Gyr	& drop~to		&$\sim$~2.3~\AA		& 6~Gyr		\\
\hline
Metallicity	&        		&		&		&        		&			&		\\
Gas		& increased		&$\sim$~0.6~dex	&$\la$200~Myr	& mildly~increased	&$\sim$~0.4~dex		& 6~Gyr		\\
in~old~stars    & unchanged             &               &               & unchanged             &                       &                \\
Average~stellar	& mildly~reduced 	& 0~to~0.15~dex	& 0~to~5~Gyr	& mildly~increased	&$\sim$~0.1~dex		& 6~Gyr		\\
\hline
Surface~brightness& decreased		&		&		& decreased		&			&		\\
FUV		& strongly~decreased	& 2-5~mag	&$\la$1~Gyr	& strongly~decreased	& 3~mag			& 6~Gyr		\\
NUV		& strongly~decreased	& 2-3~mag	&$\sim$~1~Gyr	& strongly~decreased	& 3~mag			& 6~Gyr		\\
B		& mildly~decreased	& 1~mag		&$\ga$1~Gyr	& decreased		& 2~mag			& 6~Gyr		\\
H		&$\sim$~constant	&$\la$~0.5~mag	&$\ga$1~Gyr	& decreased		& 1.5~mag		& 6~Gyr		\\
\noalign{\smallskip}
\hline
\end{tabular}
\]
}
Notes: * = Six Gyr is the lookback-time for starvation in our models. For a lookback-time $<$ 6 Gyr 
the effects induced by starvation are less important, while are more important for a lookback-time $>$ 6 Gyr.
\end{table*}


\begin{table*}
\caption{The transitional class}
\label{Tabtrans}
{\scriptsize
\[
\begin{tabular}{rcccccccccc}
\hline
\noalign{\smallskip}
VCC  & Type & Blue~Centers & Log L$_H$  &FUV-H &H$\alpha$E.W.$_{em}$&H$\beta$E.W.$_{abs}$&Log  MHI   & Age	         &V$_C$		 \\ 
     &      &              &            &      & Emission           & Absorption        &            & ($\epsilon_0$=1.2)& ($\epsilon_0$=1.2)\\
     &      &              &L$_{H\odot}$&mag   & \AA                &  \AA              & M$\odot$   &   Gyr	         &   km s$^{-1}$ \\
\noalign{\smallskip}
\hline
\noalign{\smallskip}
  21 & dS0          & bc,BC& 8.89       & 5.69 &  -                 & 4.2               & $<$7.84    &    0.50 &  59   \\
 327 & S0           &	BC & 9.34       & 4.99 &  0~(16)            & 1.1               & -          &    0.22 &  76   \\
 450 & S0 pec       &	BC & 8.86       & 5.90 &  10~(52)           & -                 & -          &    0.58 &  58   \\
 597 & S0           &	BC & 8.84       & 3.79 &  7~(36)            & -                 & -          &    0.04 &  54   \\
 710 & dS0:         &	1  & 9.00       & 3.85 &  15                & -                 &  7.95      &    0.04 &  61   \\
 764 & S0           &	1  & 8.99       & 5.63 &  -                 & -                 & -          &    0.46 &  63   \\
1065 & dE0,N        &	   & 8.34       & 5.87 &  -                 & 0.8               & -          &    0.64 &  42   \\
1175 & E5/S0        & bc,BC& 8.82       & 4.11 &  4~(157)           & -                 & -          &    0.09 &  54   \\
1389 & dE2:,N       &	   & 8.44       & 5.49 &  -                 & 1.9               & -          &    0.50 &  44   \\
1499 & E3 pec~or~S0 & bc,BC& 8.79       & 3.46 &  -                 & 5.0               & -          &    0.00 &  52   \\
1501 & dS0?         & bc,BC& 8.60       & 4.18 &  4                 & 1.1               & -          &    0.12 &  47   \\
1539 & dE0,N        &  BC  & 8.42       & 5.89 &  -                 & 1.6               & -          &    0.64 &  44   \\
1617 & d:S0 pec?    & bc,BC& 8.75       & 4.00 &  9~(7)             & 1.0               & -          &    0.08 &  52   \\
1684 & dS0:         & bc,BC& 8.76       & 5.36 &  (4)               & 4.5               & -          &    0.41 &  54   \\
1809 & S0/Sa        &	BC & 9.32       & 4.78 &  -                 & 4.9               &  7.25      &    0.18 &  75   \\
1855 & S0:          &	BC & 8.70       & 3.56 &  20~(72)           & -                 & -          &    0.01 &  50   \\

\noalign{\smallskip}
\hline
\end{tabular}
\]
}
Column 1: VCC name, from the Virgo Cluster Catalogue of Binggeli et al. (1985).\\
Column 2: Morphological classification, from Binggeli et al. (1985).\\
Column 3: Presence of a blue center in $g-i$ from Lisker et al. (2006b) (bc) and/or $NUV-i$ from this work (BC).
1 stands for galaxies not observed by the SDSS.\\
Column 4: H band luminosity, in solar units.\\
Column 5: $FUV-H$ color index, in AB magnitudes.\\
Column 6: The equivalent width of the H$\alpha$ emission line, in \AA, from (in order of preference) imaging 
(Gavazzi et al. 2006a) and integrated spectroscopy(Gavazzi et al. 2004). Nuclear (SDSS, DR5) spectroscopic data are given in parenthesis. \\
Column 7: The equivalent width of H$\beta$ underlying absorption line, in \AA, from (in order of preference) integrated (Gavazzi et al. 2004) 
or nuclear (SDSS, DR5) spectroscopy.\\
Column 8: The logarithm of the HI mass, in solar units, from Gavazzi et al. (2005b).\\
Column 9 and 10: Lookback time to the interaction and rotational velocity of the best fitting model for a ram pressure stripping event with 
$\epsilon_0$ = 1.2 M$\odot$ kpc$^{-2}$ yr$^{-1}$. Lookback times of 0 Myr are for galaxies now at the peak (maximum) of their interaction.\\ 
\end{table*}


\begin{table*}
\caption{The velocity distribution of dwarf galaxies in Virgo}
\label{Tabvel}
{\scriptsize
\[
\begin{tabular}{ccccccccccc}
\hline
\noalign{\smallskip}
		& All~Virgo&		&		&		&	& Virgo~A&		&		&	     &       \\
Class	  	& Mean  & $\sigma$ 	& Median       	& N.~of~obj.	& P	& Mean   & $\sigma$ 	& Median       	& N.~of~obj. & P \\
        	& km s$^{-1}$&km s$^{-1}$&km s$^{-1}$	&		& \%	&km s$^{-1}$&km s$^{-1}$&km s$^{-1}$	&	     &       \\
\noalign{\smallskip}
\hline
\noalign{\smallskip}
SFD		& 1362	& 790		&1324		& 151		& 4	& 1166	& 907		&1147		& 41	     & 82\\		
dE		& 1201 	& 723		&1208		& 95		& 72	& 961	& 771		& 943		& 35	     & 79\\
dE,N		& 1168	& 642		&1229		& 182		& 3	& 1120	& 701		&1222		& 89         & 12\\	
\noalign{\smallskip}
\hline
\end{tabular}
\]
}
Notes: 
Column 1: SFD stands for star forming dwarfs, low-luminosity ($L_H$$<$10$^{10}$ L$_{H \odot}$), high surface
brightness ($\mu_e(H)$$<$ 22 mag arcsec$^{-2}$) late-type galaxies.\\
Columns 6 and 11: $P$ is the probability that galaxies follow a normal distribution.\\
\end{table*}


\begin{table}
\caption{The Kolmogorov-Smirnov test for radial velocities}
\label{TabKS}
\[
\begin{tabular}{ccccccccc}
\hline
\noalign{\smallskip}
		& All~Virgo&	&	&	& Virgo~A&	&	&  	\\
Class	  	& E  	& dE 	& dE,N 	& SFD	& E  	& dE 	& dE,N 	& SFD 	\\
\noalign{\smallskip}
\hline
\noalign{\smallskip}
E		& 100	& 88	& 42	& 4	& 100	& 64	& 36	& 5 \\
dE		&  88	& 100	& 73	& 3	& 64	& 100	& 40	& 40\\
dE,N		&  42	& 73	& 100	& 0	& 36	& 40	& 100 	& 41\\	 
SFD		& 4	& 3	& 0	& 100	& 5	& 40	& 41	& 100\\
\noalign{\smallskip}
\hline
\end{tabular}
\]
SFD: Star Forming Dwarfs are all low-luminosity ($L_H$$<$10$^{10}$ L$_{H \odot}$), high surface
brightness ($\mu_e(H)$$<$ 22 mag arcsec$^{-2}$) late-type galaxies.\\
\end{table}


\begin{table}
\caption{The frequency of transient objects}
\label{Tabstat}
\[
\begin{tabular}{ccccc}
\hline
\noalign{\smallskip}
Tracer  		& Condition  	& Time~scale & fraction      &Infall~rate\\
        		&            	&     Myr    &  (N./observed)& N./Gyr    \\
\noalign{\smallskip}
\hline
\noalign{\smallskip}
HI		 	& non~def.   	& $<$150       & 16 (11/68)       & 475  \\
H$\alpha$ E.W.$_{em}$	& $>$ 2 \AA   	& $<$150       & 10 (31/319)      & 297  \\
H$\beta$ E.W.$_{abs}$	& $>$ 2.8 \AA  	& $<$500-800   & 40 (127/319)$^*$ & 274  \\
FUV-H   	 	& $<$ 5        	& $<$800       &  9 (9/102)$^{**}$& $>$50\\
\noalign{\smallskip}
\hline
\end{tabular}
\]
Notes: column 4 gives the fraction (total number/observed) of the observed early-type Virgo 
galaxies with -13.15$>$ $M_B$ $>$ -17.5 satisfying the condition, where the total number of Virgo galaxies satisfying these
criteria are 498.\\
**: 127 is the total number of galaxies with an H$\beta$ absorption line $\geq$ 2.8 \AA ~(89) or H$\beta$ in emission (38); 
if we consider also those with H$\alpha$ in emission (50), for which the H$\beta$ absorption line is generally underestimated, 
the total number of objects increases to 148 (46\%).\\
***: Given the detection limit of GALEX, the value relative to the color index $FUV-H$ is a lower limit.\\
\end{table}


\begin{table}
\caption{The lookback time to formation}
\label{Tabage}
\[
\begin{tabular}{cccc}
\hline
\noalign{\smallskip}
Model  		& Time	& 50\%  & 80\%  \\
		& Myr	& Gyr   & Gyr	\\
\noalign{\smallskip}	
\hline
\noalign{\smallskip}
Non~perturbed	&-	& 2.60	& 0.86	\\
1~crossing	& 0	& 2.72	& 0.98	\\
-		& 500	& 3.18	& 1.48	\\
-		&1000	& 3.62	& 1.96	\\
-		&1500	& 4.06	& 2.43	\\
2~crossings	&2000	& 5.98	& 4.55	\\ 
\noalign{\smallskip}
\hline
\end{tabular}
\]
Note: models are for a galaxy $V_C$ $=$ 55 km s$^{-1}$ 
for a (or two) ram pressure stripping event of efficiency $\epsilon_0$ = 1.2 M$\odot$ kpc$^{-2}$ yr$^{-1}$\\
Column 2: lookback time to the stripping event
\end{table}


\clearpage

\begin{figure*}
\epsscale{1.3} 
\plotone{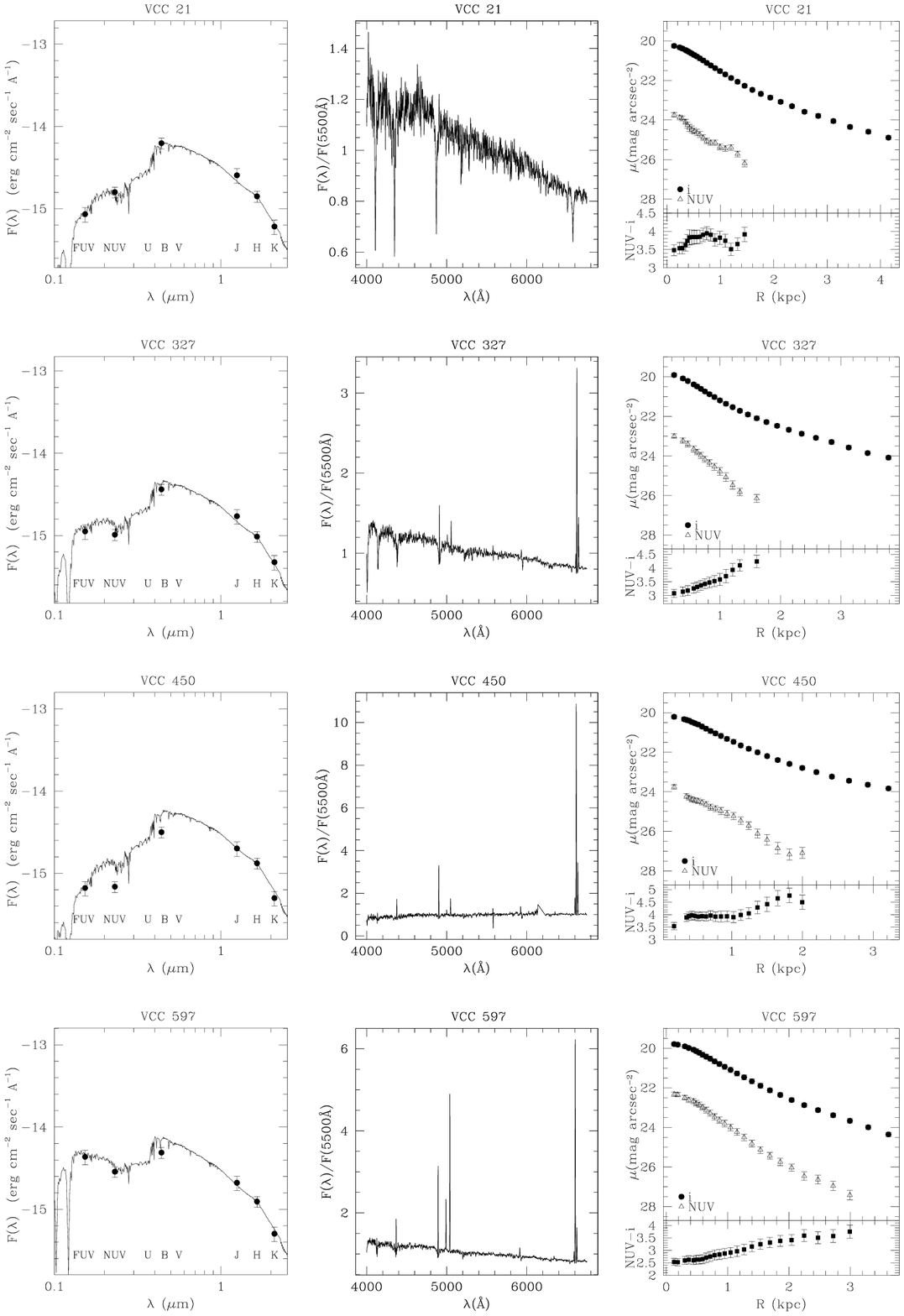}
\small{\caption{Left column: the observed (black filled dots) and model predicted (continuum line)
spectral energy distribution of the transitional galaxies listed in Table \ref{Tabtrans}  
in a ram pressure stripping scenario with $\epsilon_0$ = 1.2 M$\odot$ kpc$^{-2}$ yr$^{-1}$. The lookback time to the interaction and 
the rotational velocity of the galaxy for the model SED are listed in Table \ref{Tabtrans}. 
Central column: the integrated (form GOLDMine) or nuclear (from SDSS) visible spectrum. Only integrated
spectra have been used in the construction of the UV to near-IR SED shown in the left column. 
Right column: the $NUV-i$ radial profiles (in AB magnitudes). 
\label{sed}}}
\end{figure*}


\addtocounter{figure}{-1}
\begin{figure*}
\epsscale{1.3} 
\plotone{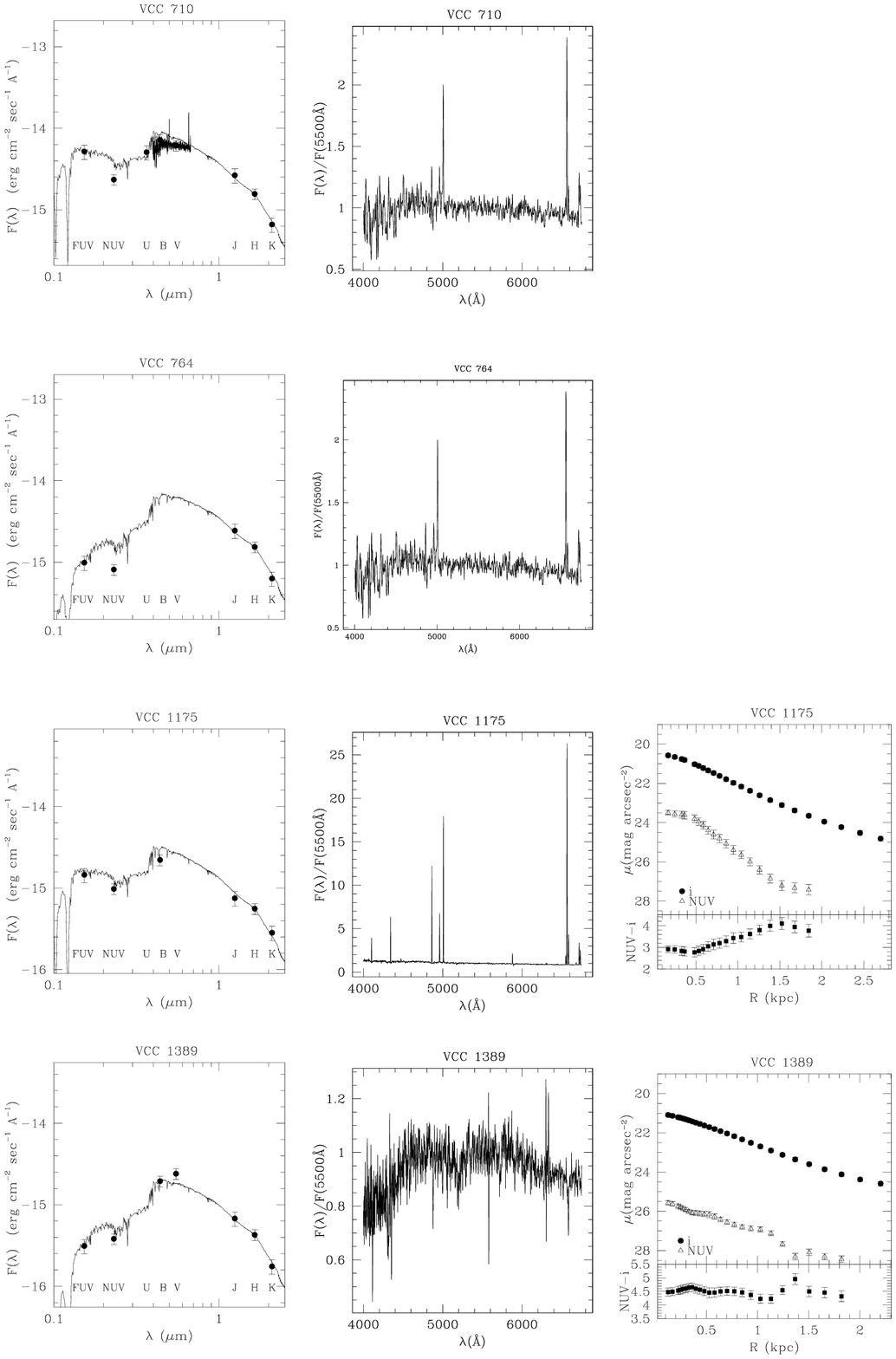}
\small{\caption{Continue. 
\label{sed}}}
\end{figure*}


\addtocounter{figure}{-1}
\begin{figure*}
\epsscale{1.3} 
\plotone{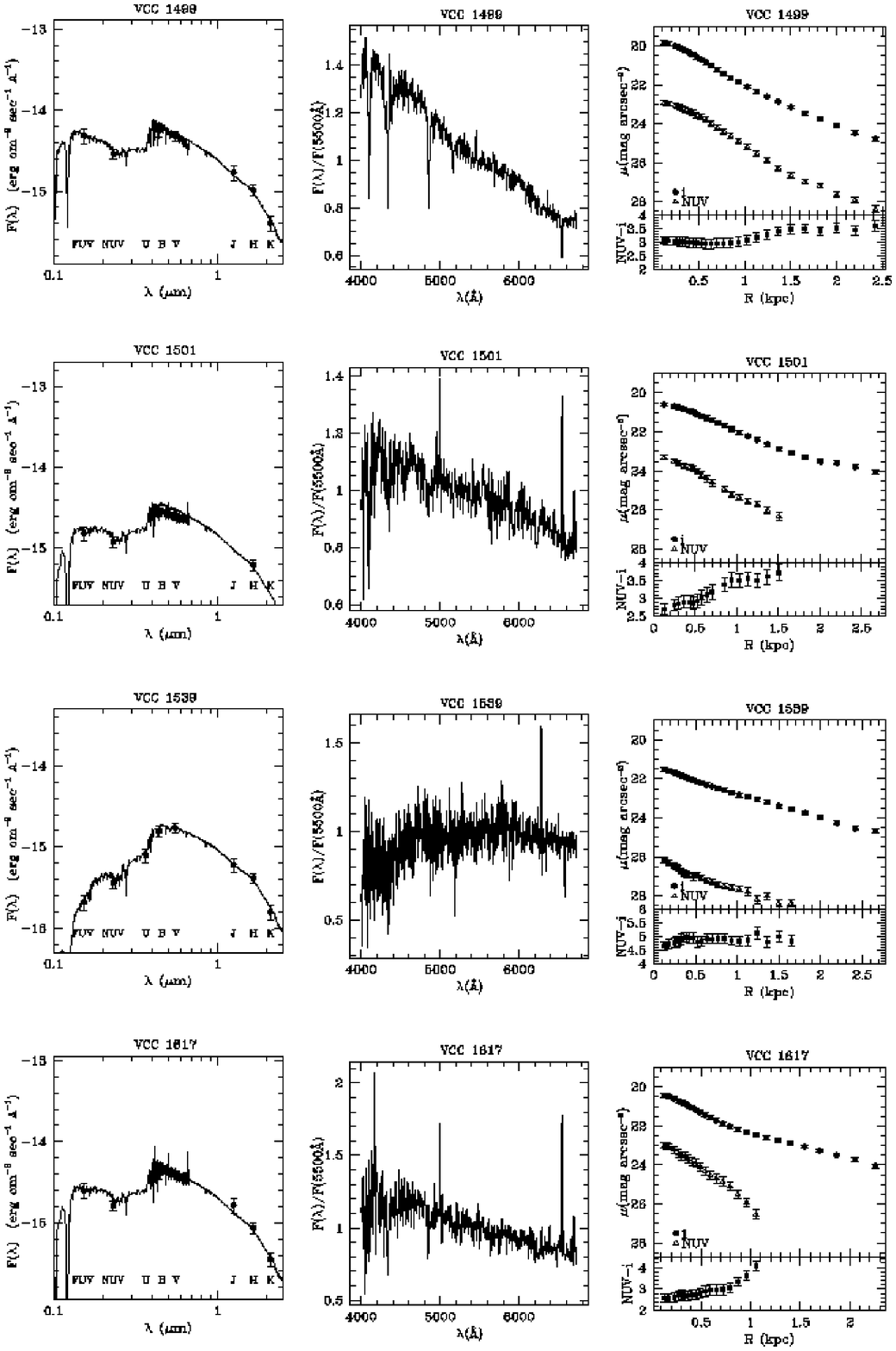}
\small{\caption{Continue. 
\label{sed}}}
\end{figure*}


\addtocounter{figure}{-1}
\begin{figure*}
\epsscale{1.3} 
\plotone{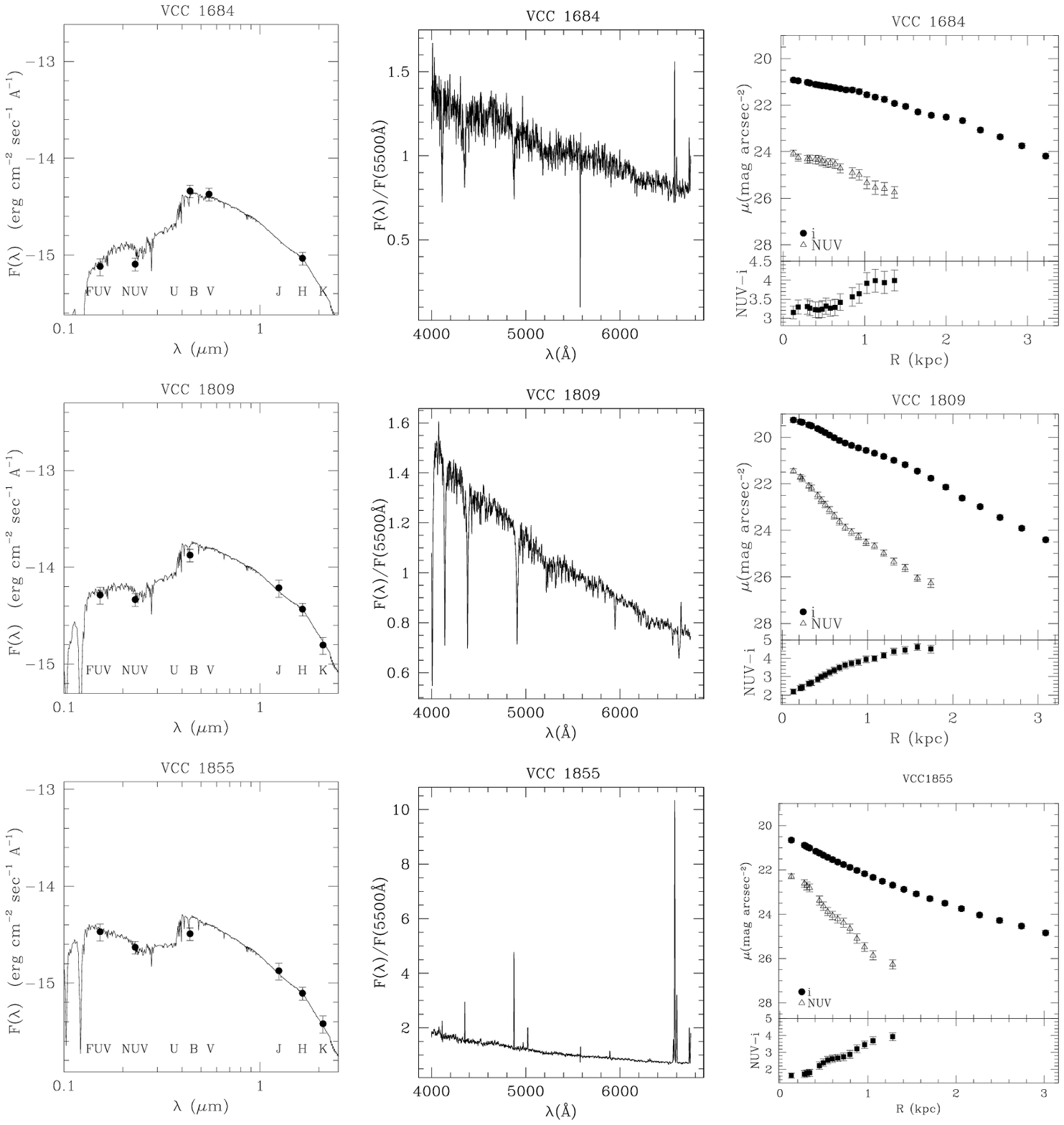}
\small{\caption{Continue. 
\label{sed}}}
\end{figure*}


\begin{figure*}
\epsscale{1.0} 
\includegraphics[angle=-90,width=15cm]{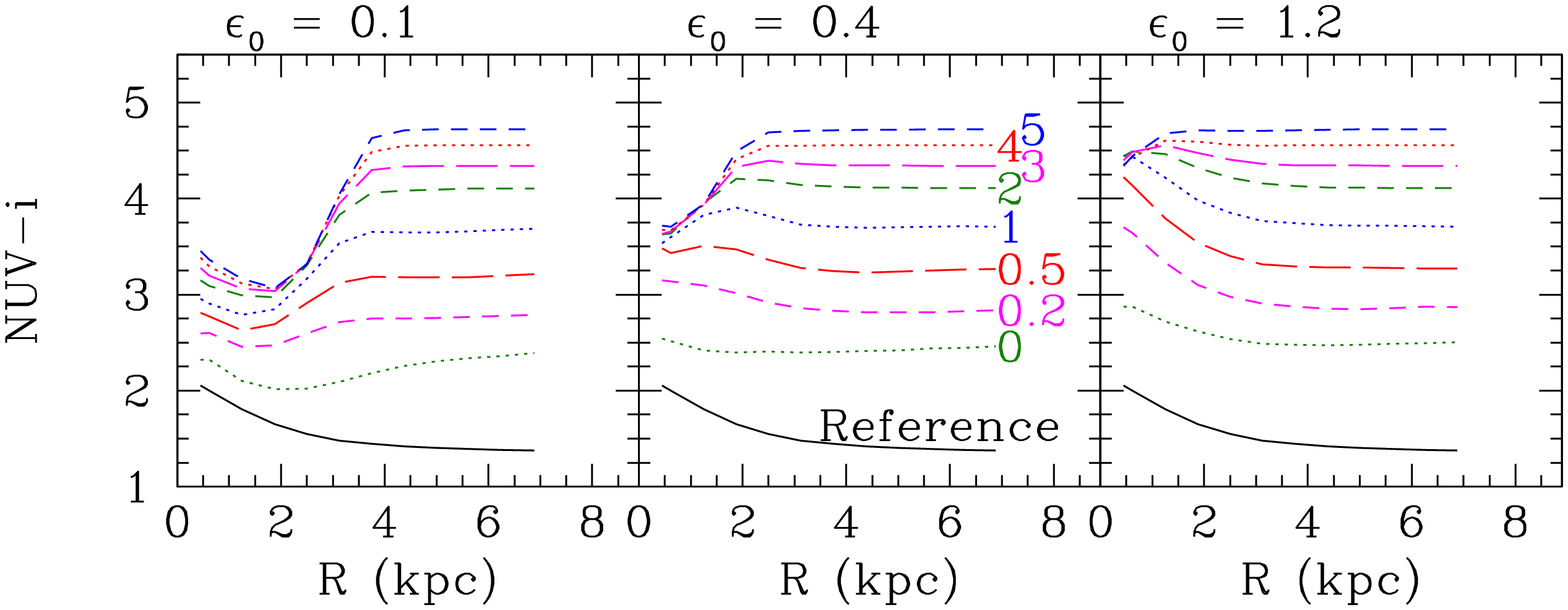}
\small{\caption{The variation of the $NUV-i$ color radial profile (in AB magnitudes) with time  
for a galaxy with rotational velocity $V_C$ = 55 km s$^{-1}$
and spin parameter $\lambda$=0.05 after a ram pressure stripping event with efficiencies
$\epsilon_0$ = 0.1 M$\odot$ kpc$^{-2}$ yr$^{-1}$ (left panel), 0.4 M$\odot$ kpc$^{-2}$ yr$^{-1}$ (central panel) 
and $\epsilon_0$ =1.2 M$\odot$ kpc$^{-2}$ yr$^{-1}$ (right panel). The lookback time to the interaction in Gyr is indicated on
the right in the central panel, from 0 (epoch when the interaction is at its maximum; 
green dotted line) to 5 Gyr (blue dashed line): 
the black continuum line is the reference unperturbed model.
\label{profiles}}}
\end{figure*}


\begin{figure*}
\epsscale{1.0} 
\plotone{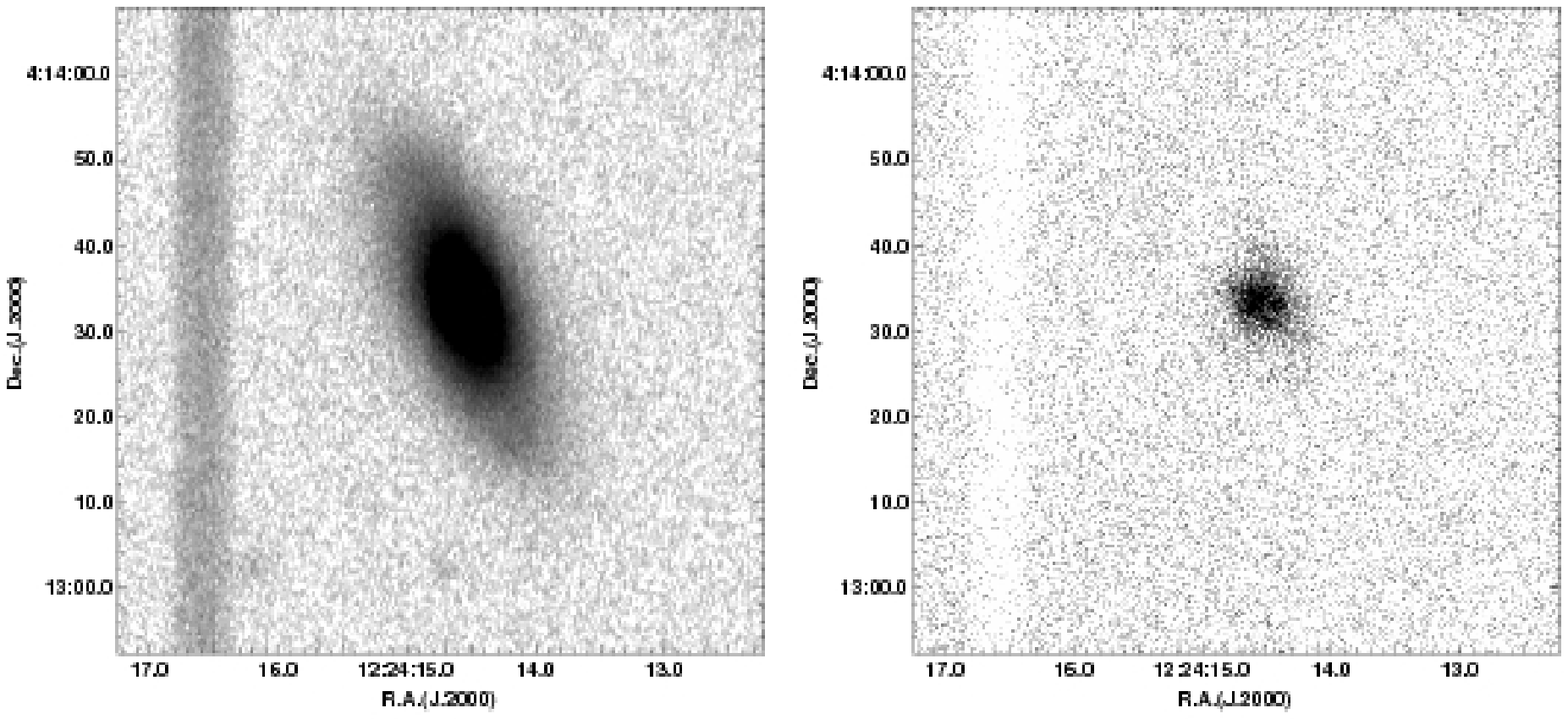}
\small{\caption{The red stellar continuum (left) and the 
H$\alpha$+[NII] narrow band images (right) of the dS0 galaxy VCC 710 from our own observations.
\label{ha}}}
\end{figure*}


\begin{figure}
\epsscale{1.0} 
\includegraphics[width=8cm,angle=0]{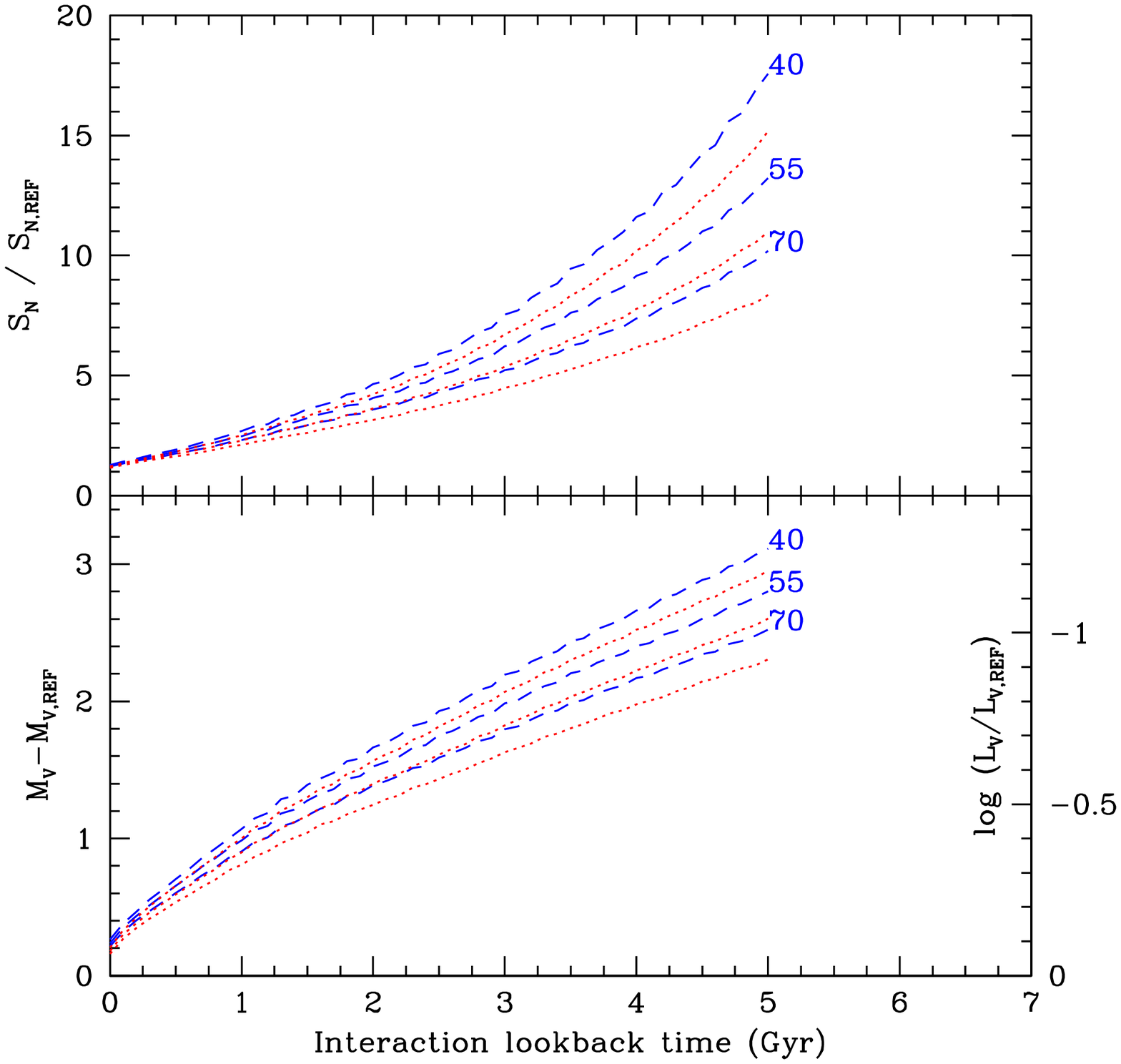}
\small{\caption{The variation of the normalized specific frequency of globular clusters (upper panel) and absolute
V band magnitude (lower panel) as a function of the lookback time to the ram
pressure stripping event for efficiencies $\epsilon_0$ = 0.4 M$\odot$
kpc$^{-2}$ yr$^{-1}$ (red dotted lines) and $\epsilon_0$ = 1.2
M$\odot$ kpc$^{-2}$ yr$^{-1}$ (short dashed lines) for galaxies
characterized by a spin parameter $\lambda$ = 0.05 and rotational
velocity $V_C$ = 40, 55 and 70 km s$^{-1}$ (from top to bottom)
respectively. The normalization is made relative to that
of a similar unperturbed galaxy.
\label{GC}}}
\end{figure}


\begin{figure}
\epsscale{1.0} \includegraphics[width=15cm,angle=0]{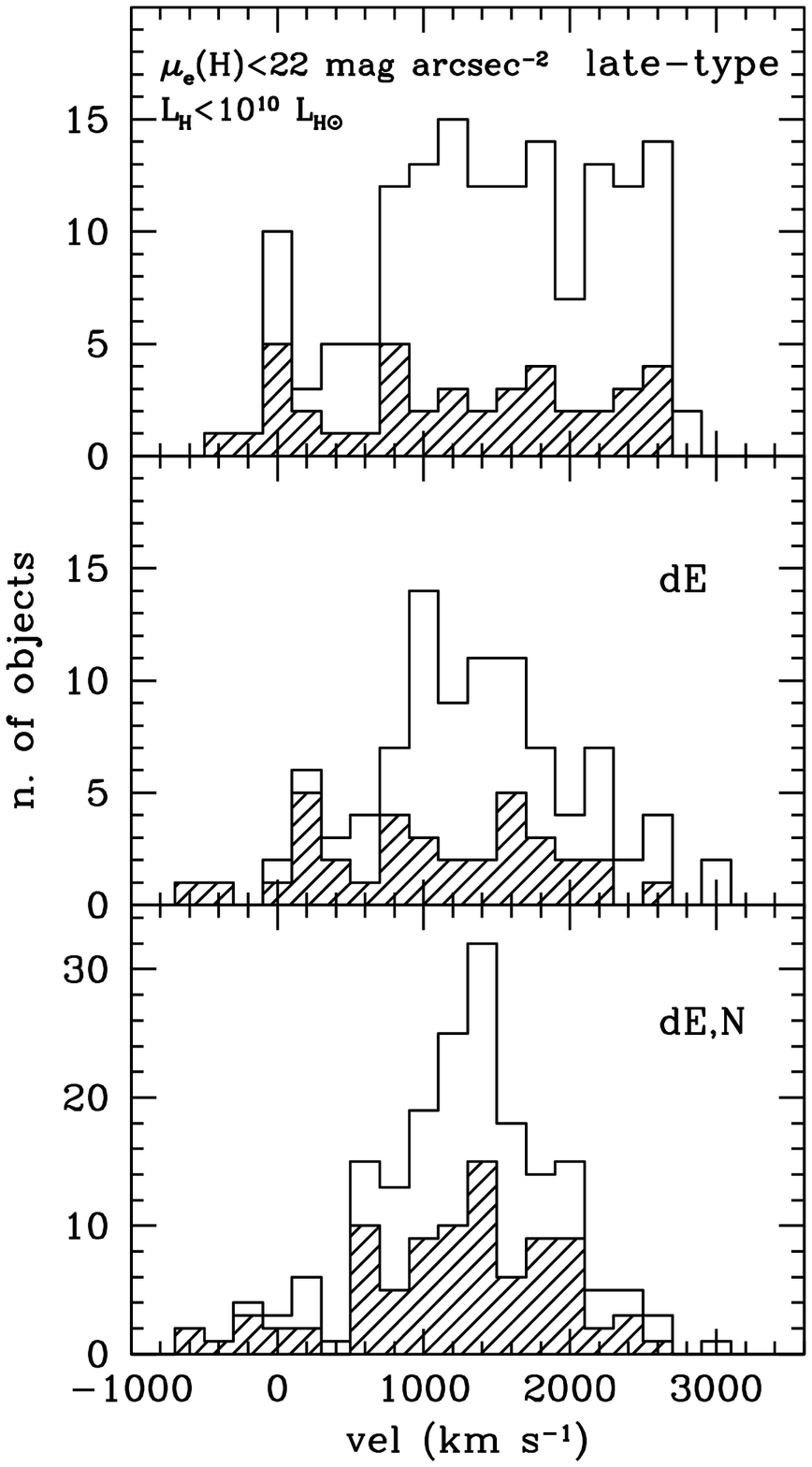}
\small{\caption{The velocity distribution of the low luminosity ($L_H$$<$ 10$^{10}$ L$_{H \odot}$), high surface brightness 
($\mu_e(H)$$<$22 mag
arcsec$^{-2}$) star forming galaxies (top), of non nucleated (center) and nucleated (bottom) dwarf ellipticals galaxies in the whole Virgo cluster
region (empty histogram) and limited to the Virgo A subcluster (hashed histogram).
\label{vel}}}
\end{figure}


\begin{figure}
\epsscale{1.0} \includegraphics[width=15cm,angle=0]{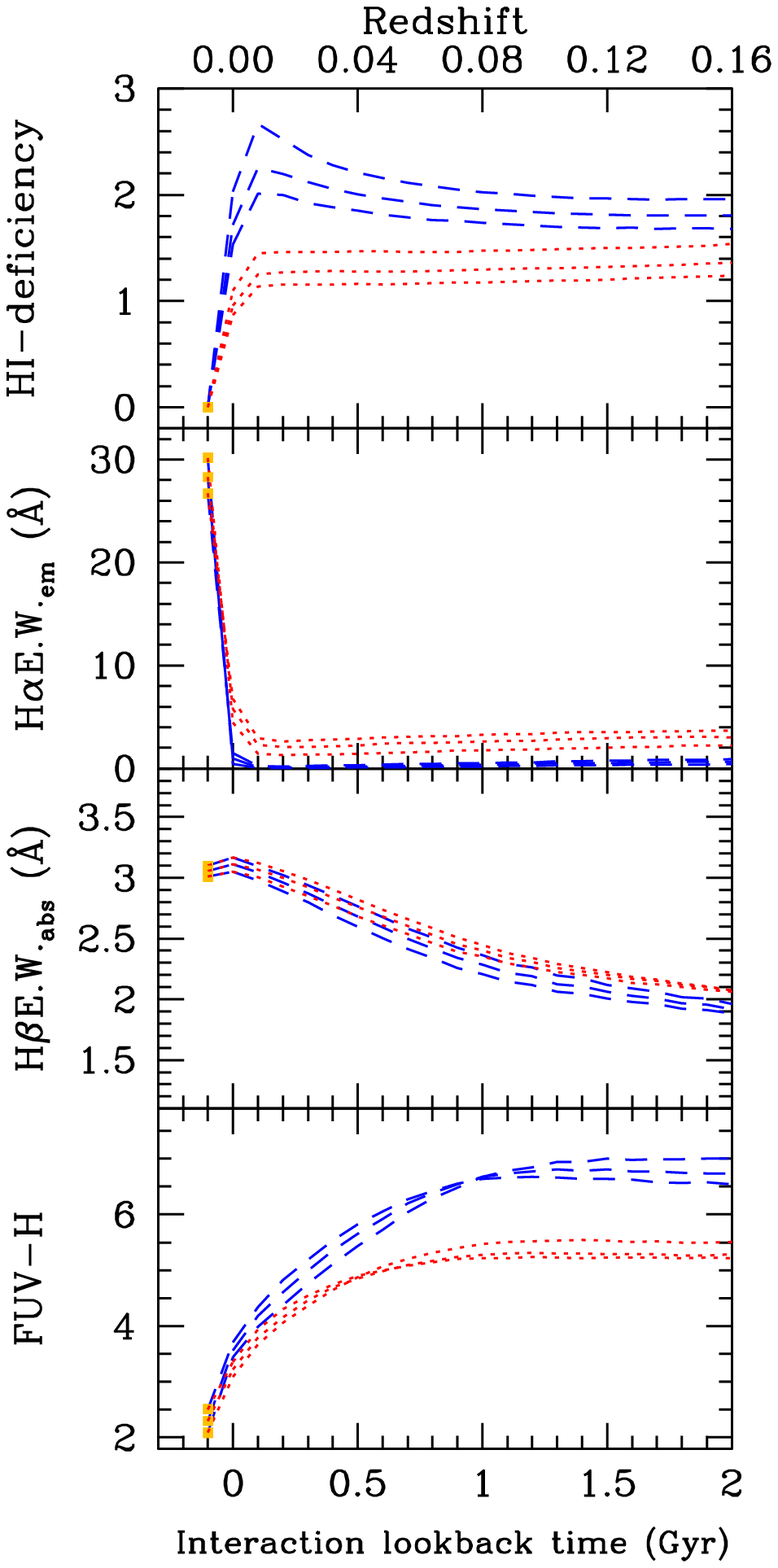}
\small{\caption{The variation of the HI-deficiency parameter (top), of the H$\alpha$E.W.$_{em}$ emission line (middle top), 
the H$\beta$E.W.$_{abs}$ absorption line (middle bottom) and the $FUV-H$
(bottom) color index for galaxies with rotational velocities of $V_C$
= 40, 55 and 70 km s$^{-1}$ respectively and spin parameter
$\lambda$=0.05 as a function of the lookback time to the interaction. The red dotted lines are for a ram pressure stripping
model with an efficiency of $\epsilon_0$ = 0.4 M$\odot$ kpc$^{-2}$
yr$^{-1}$, the blue short dashed line a ram pressure stripping model
with an efficiency of $\epsilon_0$ = 1.2 M$\odot$ kpc$^{-2}$
yr$^{-1}$. In both scenarios the efficiency of the perturbation
increases with the decrease of the rotational velocity. 
The lookback time to the interaction ($t_{rp}$) is equal to zero when the galaxy is currently passing 
through the cluster center, where ram pressure stripping is at its maximum. 
Given the extended distribution of the cluster hot gas, the galaxy-ICM
interaction in practice starts $\sim$ 100 Myr before $t_{rp}$.
For this reason, we artificially placed the values for an unperturbed model at
$t_{rp}$=-100 Myr (yellow squares).
\label{clock}}}
\end{figure}

\appendix

\section{The data}

The dataset used in the present analysis is composed of imaging and
spectroscopic data covering the whole UV to IR spectral range. The
Virgo cluster region was observed in spring 2004 as part of the All
Imaging Survey (AIS) and of the deeper Nearby Galaxy Survey (NGS) carried out
by the Galaxy Evolution Explorer (GALEX) in the two UV bands FUV ($\rm
\lambda_{eff}=1530\AA, \Delta \lambda=400\AA$) and NUV ($\rm
\lambda_{eff}=2310\AA, \Delta \lambda=1000\AA$). This public dataset
(IR 1.0 release), now available on MAST, has been combined with data
extracted from several open time observations obtained at a
similar sensitivity than the NGS. Details of the GALEX instrument and
data characteristics can be found in Martin et al$.$ (2005) and
Morrissey et al$.$ (2005).\\ 
Whenever available, we extracted fluxes
from the NGS and open time images, obtained with an average
integration time of $\sim$ 1500 sec. These data are complete to
$m_{AB}$ $\sim$ 21.5 in the NUV and FUV.  Elsewhere UV fluxes have
been extracted from the shallower AIS images ($\sim$ 70 sq. degrees), 
obtained with an average integration time of $\sim$ 100 sec,
complete to $m_{AB}$ $\sim$ 20 in both the FUV and NUV bands.  \\
B and V frames are available for most of the analyzed galaxies 
thanks to our own observations (Gavazzi \& Boselli 1996; 
Gavazzi et al. 2001; 2005a and Boselli et al. 2003).
H (1.65 $\mu$m) and K (2.1 $\mu$m) band frames have 
been obtained during a near-IR imaging survey of the Virgo cluster 
(Boselli et al. 1997b; 2000; Gavazzi et al. 2000b, 2001). 
SDSS imaging data, taken from the most recent DR5 
(release 5, Adelman-McCarthy et al. 2007), have been also used
for the determination of the radial NUV-$i$ color gradients.\\
Fluxes were obtained by integrating all
images within elliptical annuli of increasing diameter up to the
optical B band 25 mag arcsec$^{-2}$ isophotal radii. 
Independent measurements of the same
galaxies obtained in different exposures give consistent photometric
results within 10\% in the NUV and 15\% in the FUV in the AIS, and
about a factor of two better for bright (NUV $\leq$16) galaxies. 
We thus estimate that the overall uncertainty in the UV photometry is on average a factor of
$\sim$ 2 better in the NGS or open time images than in the AIS
especially for the faintest objects. Optical and near-IR data
have on average a photometric accuracy of $\sim$ 10\%.
The H and B band structural parameters effective radii $R_e$ (radius containing half of the total light) 
and surface brightnesses $\mu_e$ (mean surface brightness within $R_e$) have been measured as described 
in Gavazzi et al. (2000a). Given the poor resolution of the GALEX data (FWHM of the PSF $\sim$ 5 arcsec
for a pixel size of 1.5 arcsec), these structural 
parameters are available only for the brightest (and bluest in FUV) dwarf galaxies in Virgo.\\
Spectral line indices (H$\alpha$E.W.$_{em}$, H$\beta$E.W.$_{abs}$) are
available for a large fraction of the analyzed galaxies. Most of the
selected objects have been observed as part of a spectroscopic survey
of the Virgo cluster (Gavazzi et al. 2004) still under way. These
data have been extracted from long slit integrated spectroscopy
obtained by drifting the slit over the whole galaxy disc, as in
Kennicutt (1992), thus providing values representative of the whole
object. These are intermediate ($\lambda/\Delta\lambda \sim$ 1000)
resolution spectra in the range ($3600 - 7200$ \AA). Low resolution
spectroscopy of the central region of several galaxies is also
available from the SDSS (release 5). To minimize aperture effects,
SDSS spectroscopy has been taken only for axisymmetric quiescent
galaxies (E, dE, dS0). A few other spectroscopic data for dwarf
ellipticals have been taken from Geha et al. (2003) and Michielsen et
al. (2007). Line absorption indices (H$\beta$E.W.$_{abs}$)
have been measured in the Lick system, although their exact
calibration is still under way.\\
HI data have been taken from the compilation of Gavazzi et al. (2005b),
Conselice et al. (2003b) and (Giovanelli et al. 2007). 
The HI-deficiency parameter ($HI-def$) is defined
as the logarithmic difference between the average HI mass of a
reference sample of isolated galaxies of similar type and linear
dimension and the HI mass actually observed in individual objects:
$HI-def$ = Log$MHI_{ref}$ - Log$MHI_{obs}$. According to Haynes \&
Giovanelli (1984), Log$MHI_{ref}$ = $a$ + $b$Log$(diam)$, where $a$
and $b$ are weak functions of the Hubble type and $diam$ is the linear
diameter of the galaxy (see Gavazzi et al. 2005b).
Since we make the hypothesis that dwarf ellipticals are gas-stripped
low-luminosity, star forming objects, their HI deficiency parameter has been
determined assuming as reference isolated star forming dwarf galaxies
($a$ = 7.00 and $b$ = 1.88, from Gavazzi et al. 2005b). Galaxies with an HI-deficiency parameter
larger than 0.3 are considered as deficient in HI gas.\\
Rotational velocities for dwarf elliptical galaxies have been taken from Geha et al. (2003), van Zee et al. (2004b),
Pedraz et al. (2002), Thomas et al. (2006) and Simiens \& Prugniel (1997a, 1997b, 1998, 2002).\\
UV to near-IR imaging data have been corrected for galactic and internal extinction as described 
in Boselli et al. (2003). Internal extinction corrections have been applied only to late-type galaxies.
Owing to the high galactic latitude of Virgo, galactic extinction corrections are very small ($A_B$ $\leq$ 0.05).\\
Most of the imaging and spectroscopic data used in the present analysis are available 
in electronic format on the
GOLDMine database (http:\slash \slash goldmine.mib.infn.it; Gavazzi et al. 2003).

\section{The models}

\subsection{The multi-zone models for the chemical and spectro-photometric evolution of
unperturbed galaxies}

To trace the evolution of late-type galaxies of different luminosity, we
have used the multi-zone chemo-spectrophotometric models of Boissier
\& Prantzos (2000), updated with an empirically-determined star
formation law (Boissier et al$.$ 2003) relating the star formation
rate to the total-gas surface densities ($\Sigma_{SFR}$, $\Sigma_{gas}$):
\begin{equation}
\Sigma_{SFR}= \alpha \Sigma_{gas}^{n} V(R)/R
\end{equation}
where $V(R)$ is the rotation velocity at radius $R$. This is a
variant of the traditional ``Schmidt law'' with two parameters: the
index $n$ and an efficiency of the star formation activity $\alpha$.
The values of $\alpha$ and $n$ are taken as equal
respectively to 2.63 10$^{-3}$ and 1.48, as determined from the gaseous and
H$\alpha$ profiles of 16 nearby galaxies (Boissier et al$.$ 2003), where $\Sigma_{SFR}$ is 
expressed in M$\odot$pc$^{-2}$Gyr$^{-1}$, $\Sigma_{gas}$ in M$\odot$pc$^{-2}$, $V(R)$ in km s$^{-1}$ 
and $R$ in kpc.

The resulting models are extremely similar to those presented in Boissier \&
Prantzos (2000) and show the same global trends. We consider the star
formation law as fixed and we keep the same mass accretion (infall)
histories as in Boissier \& Prantzos (2000), based on the assumption
that before the interaction with the cluster, galaxies were ``normal''
spirals.  The free parameters in this grid of models are the spin
parameter, $\lambda$ and the rotational velocity, $V_C$.  These
two parameters are theoretical quantities, although $V_C$ should be
similar to the asymptotic value of the rotation curve at large
radii.
%
The spin parameter is a dimensionless measure of the specific angular momentum
(defined in e.g. Mo, Mao \& White 1998). Its value in spirals ranges
typically between $\sim$ 0.02 for relatively compact galaxies to
$\sim$ 0.09 for low surface brightness galaxies (Boissier \& Prantzos
2000). The models of Boissier \& Prantzos (2000) contains scaling
relationships (the total mass varies as $V_C^3$, the scale-length as
$\lambda \times V_C$). Star formation histories depend on the infall
timescales, which are a function of $V_C$ in these models, so that 
roughly speaking,
$V_C$ controls the stellar mass accumulated during
the history of the galaxy, and $\lambda$ its radial distribution.\\
In order to avoid the multiplication of parameters,
we decided to fix the spin parameter to an average value of 0.05 and investigate global
trends linked to the mass -or the velocity of galaxies- or to the interaction itself.
Other values of the spin parameter should mostly create some scatter around these trends.
%
From e.g. Boissier \& Prantzos (2000), it can be seen that trends 
are mostly driven by mass (see their Figure 8). Various spin parameters create
a modest dispersion with respect to the average one (e.g. about a few tenths of
magnitude for the B-K colour index, a few tenths of dex for stellar and gaseous
masses...). The only quantities which are clearly much more affected by the
spin parameter than the velocity are the surface brightnesses.
The central surface brightness in the B band (or $r$ band, see Boissier 2000) can
change by a few mag arcsec$^{-2}$ for a given circular velocity when using
various realistic spin parameters.

To sample the whole dynamic range covered by dwarf galaxies we focused our 
analysis to the velocity range 40-100 km s$^{-1}$, 
although we tested several other models for larger velocities (for
comparison with bright spirals). We remind that the model, which has a resolution of $\sim$ 1 kpc, does not
include any bulge or nuclear component, 
corresponding well to pure exponential disks such as these dwarf 
star forming galaxies (Gavazzi et al. 2000a). 
On the other hand, the models were developed and extensively tested for
relative massive galaxies (rotation velocities above 80 km s$^{-1}$).
In this work, we are interested mostly in the 40-100  km s$^{-1}$ range, and
low velocity models were simply extrapolated from the massive ones.
By this, we mean that we assumed that the star formation law, Initial
Mass Function and other ingredients used for massive galaxies also
apply to the low mass case. Especially, the models include a dependence of infall time
scales on the circular velocity (more massive galaxies form the bulk
of their stars earlier than lower mass galaxies). This assumption
was found to be in agreement with observations in normal spirals,
above $\sim$ 80 km/s (Boissier \& Prantzos, 2000), and is here generalized
to lower masses.
%
%
These considerations call for some caution in the interpretation of
our results. Especially, low mass galaxies are very small and the
models do not take into account the likely radial mixing on scales
smaller than $\sim$ 1 kpc, making them extremely uncertain with
respect to e.g. metallicity gradients. Nevertheless, we show in this paper
that our extrapolated models reproduce reasonably well the properties
of normal late type low velocity galaxies (magnitudes, colors...).
In addition, errors due to model uncertainties are likely to be
systematic effects: they should affect less differences between two
models than absolute values. Thus, we believe that the differences
obtained when ram-pressure is introduced with respect to the
unperturbed case can be trusted.


%
The models of Boissier \&
Prantzos (2000) provide total luminosities and colors in all bands as well
as the total gas content. 
They do not compute the nebular emission,
but we estimated the H$\alpha$ emission by using the number of
ionizing photons predicted by Version 5 of STARBURST 99 (Vazquez \&
Leitherer 2005) for a single generation of stars distributed on the
Kroupa et al. (1993) initial mass function (as used in our models),
convolving it with our star formation history, and converting the
result into our  H$\alpha$ flux following Appendix A of Gavazzi et
al. (2002a). \\
The models also produce low resolution theoretical 
spectra from which are used to compute the Lick H$\beta$
index. The model spectral resolution is actually quite poor 
(20 \AA~in the visible) but we
checked that the evolution of this index in the case of a burst is
consistent with the one presented in e.g. Tantalo \& Chiosi (2004).  \\
The evolution of the metallicity is computed in the models taking
into account the yields, finite lifetimes of stars, infall of pristine
gas (Boissier \& Prantzos 1999). The spectra are computed taking
into account this metallicity and the star formation history.


%
Given the agreement between model predictions and observations for different
samples of unperturbed galaxies (Boissier \& Prantzos 1999; 2000; Prantzos \& Boissier 2000; 
Boissier et al. 2001), we will consider them as the
reference models for the unperturbed case. As for NGC 4569 (Boselli et al. 2006)
we study two different interaction scenarios, starvation and ram-pressure stripping.

\subsection{The starvation scenario}
 
In the starvation scenario (Larson et al. 1980, Balogh et al. 2000,
Treu et al. 2003), the cluster acts on large scales by removing any
extended gaseous halo surrounding the galaxy, preventing further infall
of such gas onto the disk. The galaxy then become anemic simply because it 
exhausts the gas reservoir through on-going star formation.\\
Infall is a necessary assumption in models of the chemical
evolution of the Milky Way to account for the G-dwarf metallicity
distribution (Tinsley 1980) and is supported by some chemo-dynamical
models (Samland et al. 1997). As the disk galaxy models were obtained
through a generalization of the Milky Way model, infall is present in
all our models. It is a schematic way to describe the
growth of any galaxy from a proto-galactic clump in the distant past to
a present-day galaxy. Infall time scales in the models were
chosen to reproduce the properties of present day normal
galaxies (Boissier \& Prantzos 2000, Boissier et al. 2001). 
This includes a dependency on the 
rotational velocity such that infall in low mass galaxies is shifted towards
later time: this mimics the fact that massive galaxies are formed early on
and that low mass galaxies are still forming at the current time, a well known
fact (Gavazzi et al. 1996, 2002; Boselli et al. 2001) nowadays called ``downsizing''. Because of this effect,
low mass galaxies, in the absence of interaction, are accreting large amounts of gas
and are still actively forming stars. In the event of starvation, the galaxy 
stops accreting gas, and continue forming stars from the reservoir already present
in the disk. This proceed at a lower rate than in the unperturbed case, with a 
declining activity in time as the consumed gas is not replaced with fresh gas.
Stopping infall (in order to mimic starvation)
at a given time is straightforward to include in the model. We shall
call $t_s$ the elapsed time since the infall termination (look back time).

\subsection{The ram pressure stripping scenario}

In addition to the starvation scenario we can study the effect of
ram-pressure gas stripping. As in Boselli et al. (2006), we adopt the
plausible scenario of Vollmer et al. (2001) explicitly tailored to
Virgo: i.e., the galaxies being modeled have crossed the dense IGM
only once (in its simplest version), on elliptical orbits. The ram
pressure exerted by the IGM on the galaxy ISM varies in time following
a Gaussian profile, whose peak at ($t$=$t_{rp}$) is when the galaxy is
crossing the dense cluster core at high velocity ($t$ and $t_{rp}$ are
look-back times, where the present epoch corresponds to $t$=0). The
Gaussian has a width $\Delta t$ = 9 $\times$ 10$^7$ years (see
Fig$.$~3 of Vollmer et al. 2001). We make the hypothesis that the gas
at each radius is removed at a rate that is directly proportional to
the galaxy gas column density $\Sigma_{gas}$ and inversely
proportional to the potential of the galaxy, measured by the total
(baryonic) local density $\Sigma_{potential}$ (provided by the model).
The gas-loss rate adopted is then equal to $\epsilon$
$\frac{\Sigma_{gas}}{\Sigma{potential}}$, with the efficiency
$\epsilon$ following a Gaussian having a maximum $\epsilon_0$ at the
time $t_{rp}$, chosen to mimic the variation of the ram pressure
suggested by Vollmer et al. (2001). This very simple, but
physically-motivated prescription should allow us to model the gas
removal from galaxies using only two free parameters ($t_{rp}$ and
$\epsilon_0$) to age-date and measure the magnitude of this effect. To
further constrain the models we assume two different values of
$\epsilon_0$: the first is $\epsilon_0$ = 1.2 M$\odot$ kpc$^{-2}$
yr$^{-1}$, the value best reproducing the radial profiles of NGC 4569
(Boselli et al., 2006), the second one is one third of it,
$\epsilon_0$ = 0.4 M$\odot$ kpc$^{-2}$ yr$^{-1}$. This second value is
suggested by the following considerations: besides the potential of
the galaxy, which is an intrinsic property of each object, the
efficiency of ram-pressure ($P$) stripping depends on the density of
the cluster intergalactic medium ($\rho_{IGM}$), on the crossing
velocity of the galaxy within the cluster ($V_{gal}$) and on its
orientation with respect to its motion (Boselli \& Gavazzi 2006): $P$
= $\rho_{IGM}$ $V_{gal}^2$.  Statistically speaking, the line of sight
velocity dispersion of star forming galaxies in the cluster is 1150 km
s$^{-1}$, and drops to 766 km s$^{-1}$ if limited to low luminosity
star forming objects in the Virgo A subgroup from where the X ray
emission comes from.  These values can be compared to the line of
sight velocity of NGC 4569 with respect to the cluster center, which
is 1374 km s$^{-1}$.  To test other realistic models we thus decided
to use an $\epsilon_0$ lower by a factor of $\sim$ (1374/766)$^2$ = 3.2
than in NGC 4569.  $\epsilon_0$ = 0.4 M$\odot$ kpc$^{-2}$
yr$^{-1}$ can thus be considered as an average value of relatively
weak interactions while $\epsilon_0$ = 1.2 M$\odot$ kpc$^{-2}$
yr$^{-1}$ is a realistic value representative of more extreme
interactions. As we discuss in the main body of the paper, $\epsilon_0$ = 0.4
M$\odot$ kpc$^{-2}$ yr$^{-1}$ leads to HI-deficiency parameters of
$\sim$ 0.8 for massive galaxies, consistent with the average
HI-deficiency observed within one virial radius in Virgo, while
$\epsilon_0$ = 1.2 M$\odot$ kpc$^{-2}$ yr$^{-1}$ to HI-deficiencies of
$\sim$ 1.5, close to the highest values found in Virgo (Gavazzi et
al. 2005b; Boselli \& Gavazzi 2006). In massive galaxies HI
deficiencies as high as $\sim$ 2 have been observed (Gavazzi et
al. 2005), and are probably the result of stripping events more
efficient than those simulated.  These extreme HI deficiencies can
also be the result of subsequent crossings of the cluster center, that
should happen on average every $\sim$ 1.7 Gyr (Boselli \& Gavazzi
2006). Multiple crossing can be reproduced by our models just
considering that the average crossing time of star forming galaxies in
the Virgo cluster is 1.7 Gyr (Boselli \& Gavazzi 2006).\\ 
We make the further assumption that no extra star formation is induced during the
interaction. This assumption is reasonable since ram pressure stripping 
models of Fujita (1998) and Fujita \& Nagashima (1999) indicate that 
on short time scales ($\sim$ 10$^8$ yr) the star formation
activity of galaxies can increase by up to a factor of 70\% at most 
in high density, rich clusters such as Coma, but it decreases to values lower than before the interaction
whenever the density of the IGM is relatively low as in the core 
of Virgo. Indeed we do not have any evidence of a statistically significant increase of the star formation
activity in galaxies that recently underwent a ram-pressure
stripping event (Iglesias-Paramo et al. 2004).
Given the rapidity of the gas stripping process (150 Myr, see sect. 5.2) in these dwarf 
systems, a possible mild increase of their star formation activity does not have enough
time to produce significant modification in their 
spectrophotometric and structural properties.

\references


\reference{}Adami, C., Biviano, A., Durret, F., Mazure, A., 2005, A\&A, 443, 17

\reference{}Adelman-McCarthy J., et al., 2007, ApJS, submitted

\reference{}Auld, R., Minchin, R., Davies, J., et al., 2006, MNRAS, 371, 1617

\reference{}Balogh, M.L., Navarro, J.F., \& Morris, S.L., 2000, ApJ, 540, 113


\reference{}Bender, R., Burstein, D., Faber, S., 1992, ApJ, 399, 462


\reference{}Binggeli, B., Sandage, A., Tammann, G., 1985, AJ, 90, 1681

\reference{}Binggeli, B., Sandage, A., Tammann, G., 1988, ARA\&A, 26, 509

\reference{}Binggeli, B., Tarenghi, M., Sandage, A., 1990, A\&A, 228, 42


\reference{}Binggeli, B., Popescu, C., Tammann, G., 1993, A\&AS, 98, 275

\reference{}Binggeli, B., Popescu, C., 1995, A\&A, 298, 63

\reference{}

\reference{}Blanton, M. R., Lupton, R., Schleghel, D., Strauss, M., Brinkmann, J., Fukugita, M., Loveday, J., 2005, ApJ, 631, 508 

\reference{}Boissier, S., \&  Prantzos, N.\ 1999, MNRAS, 307, 857 

\reference{}Boissier, S. \ 2000, Ph.D.~Thesis,  

\reference{}Boissier, S. \& Prantzos, N., 2000, MNRAS, 312, 398 

\reference{}Boissier, S., Boselli, A., Prantzos, N., \& Gavazzi, G.\ 2001, MNRAS, 321, 733 

\reference{}Boissier, S., Prantzos, N., Boselli, A. \& Gavazzi, G., 2003, MNRAS, 346, 1215 



\reference{}Boselli, A., 1994, A\&A, 292, 1

\reference{}Boselli, A. \& Gavazzi, G., 2006, PASP, 118, 517

\reference{}Boselli, A., Gavazzi, G., Combes, F., Lequeux, J., Casoli, F., 1994, A\&A, 285, 69

\reference{}Boselli, A., Gavazzi, G., Lequeux, J., Buat, V., Casoli, F., Dickey, J., Donas, J., 1997a, A\&A, 327, 522

\reference{}Boselli, A., Tuffs, R., Gavazzi, G., Hippelein, H. \& Pierini, D., 1997b, A\&AS, 121, 507

\reference{}Boselli, A., Gavazzi, G., Franzetti, P., Pierini, D., Scodeggio, M., 2000, A\&AS, 142, 73

\reference{}Boselli, A., Gavazzi, G., Donas, J. \& Scodeggio, M., 2001, AJ, 121, 753

\reference{}Boselli, A., Lequeux, J. \& Gavazzi, G., 2002, A\&A, 384, 33

\reference{}Boselli, A., Gavazzi, G. \& Sanvito, G., 2003, A\&A, 402, 37


\reference{}Boselli, A., Cortese, L., Deharveng, JM., et al., 2005a, ApJ, 629, L29

\reference{}Boselli, A., Boissier, S., Cortese, L., et al., 2005b, ApJ, 623, L13

\reference{}Boselli, A., Boissier, S., Cortese, L., Gil de Paz, A., Seibert, M., Madore, B.F., Buat, V., Martin, D.C., 2006,
ApJ, 651, 811

\reference{}Bothun, G., Mould, J., Caldwell, N., MacGillivray, H., 1986, AJ, 92, 1007

\reference{}Bower, R.G., Lucey, J.R., Ellis, R.S., 1992, MNRAS, 254, 601

\reference{}Buat, V., Xu, K., 1996, A\&A, 306, 61


\reference{}Bullock, J., Kravtsov, A., Weinberg, D., 2000, ApJ, 539, 517

\reference{}Cassata, P., Guzzo, L., Franceschini, A., et al., 2007, ApJS, in press (astroph/0701483)


\reference{}Catinella, B., Giovanelli, R., Haynes, M., 2006, AJ, 640, 751




\reference{}Cole, S., Aragon-Salamanca, A., Frenk, C., Navarro, J., Zepf, S., 1994, MNRAS, 271, 781

\reference{}Conselice, C., 2002, ApJ, 573, L5

\reference{}Conselice, C., Gallagher, J., Wyse, R., 2001, ApJ, 559, 791

\reference{}Conselice, C., Gallagher, J., Wyse, R., 2003a, AJ, 125, 66

\reference{}Conselice, C., O'Neil, K., Gallagher, J., Wyse, R., 2003b, ApJ, 591, 167



\reference{}C\^ot\'e P., Piatek S., Ferrarese L., et al., 2006, ApJS, 165, 57




\reference{}Davies, J., Phillipps, S., 1988, MNRAS, 233, 553

\reference{}Dekel, A., Silk, J., 1986, ApJ, 303, 39

\reference{}De Lucia, G., Poggianti, B., Aragon-Salamanca, A., et al., 2004, ApJ, 610, L77

\reference{}De Lucia, G., Springel, V., White, S., Croton, D., Kauffmann, G., 2006, MNRAS, 366, 499

\reference{}De Lucia, G., Poggianti, B., Aragon-Salamanca, A., et al., 2007, MNRAS, 374, 809

\reference{}Dressler, A., 1980, ApJ, 236, 351


\reference{}Dressler, A., in "Clusters of Galaxies: Probes of Cosmological Structure and Galaxy Evolution", Cambridge University Press, ed. by Mulchaey et al., 2004, p. 207

\reference{}Duc, P.-A., Cayatte, V., Balkowski, C., Thuan, T.X., Papaderos, P., van Driel, W., 2001, A\&A, 369, 763 

\reference{}Elmegreen, D., Elmegreen, B., Frogel, J., Eskridge, P., Pogge, R., Gallager, A., Iams, J., 2002, AJ, 124, 777


\reference{}Ferguson, H., Binggeli, B., 1994, A\&ARev, 6, 67

\reference{}Ferrara, A., Tolstoy, E., 2000, MNRAS, 313, 291

\reference{}Franx, M., in "Clusters of Galaxies: Probes of Cosmological Structure and Galaxy Evolution", Cambridge University Press, ed. by
Mulchaey et al., 2004, p. 197

\reference{}Fuchs, B., \& vn Linden, S., 1998, MNRAS, 294, 513

\reference{}Fujita, Y., 1998, ApJ, 509, 587


\reference{}Fujita, Y., Nagashima, M., 1999, ApJ, 516, 619

\reference{}Gavazzi, G., Boselli, A., 1996, Astro. Lett. and Communications, 35, 1

\reference{}Gavazzi, G., Boselli, A., 1999, A\&A, 343, 93

\reference{}Gavazzi, G., Boselli, A., Kennicutt, R., 1991, AJ, 101, 1207

\reference{}Gavazzi, G., Pierini, D. \& Boselli, A., 1996, A\&A, 312, 397  

\reference{}Gavazzi, G., Catinella, B., Carrasco, L., Boselli, A., Contursi, A., 1998, AJ, 115, 1745

\reference{}Gavazzi, G., Boselli, A., Scodeggio, M., Pierini, D., Belsole, E., 1999, MNRAS, 304, 595

\reference{}Gavazzi, G., Franzetti, P., Scodeggio, M., Boselli, A. \& Pierini, D., 2000a, A\&A, 361, 863

\reference{}Gavazzi, G., Franzetti, P., Scodeggio, M., Boselli, A., Pierini, D., Baffa, C., Lisi, F., Hunt, L., 2000b, A\&AS,
142, 65

\reference{}Gavazzi, G., Zibetti, S., Boselli, A., Franzetti, P., Scodeggio, M., Martocchi, S., 2001, A\&A, 372, 29 


\reference{}Gavazzi, G., Boselli, A., Pedotti, P., Gallazzi, A. \& Carrasco, L., 2002a, A\&A, 396, 449

\reference{}Gavazzi, G., Bonfanti, C., Sanvito, G., Boselli, A., \& Scodeggio, M., 2002b, ApJ, 576, 135 

\reference{}Gavazzi, G., Boselli, A., Donati, A., Franzetti, P. \& Scodeggio, M., 2003, A\&A, 400, 451

\reference{}Gavazzi, G., Zaccardo, A., Sanvito, G., Boselli, A. \& Bonfanti, C., 2004, A\&A, 417, 499

\reference{}Gavazzi, G., Donati, A., Cuccati, O., Sabatini, S., Boselli, A., Davies, J., Zibetti, S., 2005a, A\&A, 430, 411 

\reference{}Gavazzi, G., Boselli, A., van Driel, W., O'Neil, K., 2005b, A\&A, 429, 439

\reference{}Gavazzi, G., Boselli, A., Cortese, L., Arosio, I., Gallazzi, A., Pedotti, P., Carrasco, L., 2006a, A\&A, 443, 839

\reference{}Gavazzi, G., O'Neil, K., Boselli, A., van Driel, W., 2006b, A\&A, 449, 929

\reference{}Geha, M., Guhathakurta, P., van der Marel, R.P., 2003, AJ, 126, 1794


\reference{}Gil de Paz, A., Boissier, S., Madore, B., et al., 2007, ApJS, in press 

\reference{}Giovanelli, R., Haynes, M., Kent, B., et al., 2005, AJ, 130, 2598

\reference{}Giovanelli, R., Haynes, M., Kent, B., et al., 2007, AJ, submitted


\reference{}Graham, A., Guzman, R., 2003, AJ, 125, 2936

\reference{}Graham, A., Jerjen, H., Guzman, R., 2003, AJ, 126, 1787

\reference{}Grebel, E., 1999, IAUS, 192, 17

\reference{}Grebel, E., Gallagher, J., Harbeck, D., 2003, AJ, 125, 1926

\reference{}Gunn, J.E., \& Gott, J.R.I., 1972, ApJ, 176, 1



\reference{}Haynes, M., \& Giovanelli, R., 1984, AJ, 89, 758





\reference{}Iglesias-Paramo, J., Boselli, A., Gavazzi, G., Zaccardo, A., 2004, A\&A, 421, 887

\reference{}Kauffmann, G., White, S., Guiderdoni, B., 1993, MNRAS, 264, 201




\reference{}Kennicutt, R., 1992, ApJ, 388, 310

\reference{}Kennicutt, R., 1998, ARA\&A, 36, 189

\reference{}Klypin, A., Kravtsov, A., Valenzuela, O., Prada, F., 1999, ApJ, 522, 82





\reference{}Kroupa, P., Tout, C.~A., \& Gilmore, G.\ 1993, MNRAS, 262, 545 


\reference{}Larson, R., Tinsley, B. \& Caldwell, N., 1980, ApJ, 237, 692


\reference{}Lin, D., Faber, S., 1983, ApJ, 266, L21

\reference{}Lisker, T., Grebel, E., Binggeli, B., 2006a, AJ, 132, 497

\reference{}Lisker, T., Glatt, K., Westera, P., Grebel, E., 2006b, AJ, 132, 2432

\reference{}Lisker, T., Grebel, E., Binggeli, B., Glatt, K., 2007, ApJ, in press (astroph/0701429)

\reference{}Lotz J., Miller B., Ferguson H., 2004, ApJ, 613, 262 

\reference{}Mac Low, M., Ferrara, A., 1999, ApJ, 513, 142

\reference{}Marcolini, A., Brighenti, F., D'Ercole, A., 2003, MNRAS, 345, 1329


\reference{}Martin, C., Fanson, J., Schiminovich, D., et al., 2005, ApJ, 619, L1

\reference{}Mastropietro, C., Moore, B., Mayer, L., Debattista, V., Piffaretti, R., Stadel, J., 2005, MNRAS,
364, 607

\reference{}Mateo, M., 1998, ARA\&A, 36, 435

\reference{}Mayer, L., Governato, F., Colpi, M., Moore, B., Quinn, T., Wadsley, J., Stadel, J., Lake, G., 2001a,
ApJ, 547, L123

\reference{}Mayer, L., Governato, F., Colpi, M., Moore, B., Quinn, T., Wadsley, J., Stadel, J., Lake, G., 2001b,
ApJ, 559, 754


\reference{}Michielsen, D., Boselli, A., Conselice, C., et al., 2007, MNRAS, submitted

\reference{}Miller, B., Lotz, J., Ferguson, H., Stiavelli, M., Whitmore, B., 1998, ApJ, 508, L133

\reference{}Mo, H.~J., Mao, S., \&  White, S.~D.~M.\ 1998, MNRAS, 295, 319 


\reference{}Moore, B., Lake, G., katz, N., 1998, ApJ, 495, 139

\reference{}Mori, M., \& Burkert, A., 2000, ApJ, 538, 559

\reference{}Morrissey, P., Schiminovich, D., Barlow, T., et al., 2005, ApJ, 619, L7

\reference{}Murakami, I., \& Babul, A., 1999, MNRAS, 309, 161

\reference{}Nagashima, M., Yahgi, H., Enoki, M., Yoshii, Y., Gouda, N., 2005, ApJ, 634, 26

\reference{}Nelan, J., Smith, R., Hudson, M., Wegner, G., Lucey, J., Moore, S., Quinney, S., \& Suntzeff, N., 2005, ApJ, 632, 137


\reference{}Nolan, L.A., 2004, in "Clusters of Galaxies: Probes of Cosmological Structure and Galaxy Evolution", Cambridge University
Press, ed. by Mulchaey et al., 2004, p. 38


\reference{}Pasquali A., Larsen S., Ferreras I., Gnedin O., Malhotra S., Rhoads J., Pirzkal N., Walsh J., 2005, AJ, 129, 148

\reference{}Pedraz, S., Gorgas, J., Cardiel, N., S\'anchez-Bl\'asquez, P., Guzm\'an, R., 2002, MNRAS, 332, L59

\reference{}Poggianti, B.M., \& Barbaro, G., 1997, A\&A, 325, 1025

\reference{}Poggianti, B.M., Bridges, T.J., Mobasher, B., et al., 2001a, ApJ, 562, 689

\reference{}Poggianti, B.M., Bridges, T.J., Carter, D., et al., 2001b, ApJ, 563, 118


\reference{}Popesso, P., Biviano, A., B\"oringer, H., Romaniello, M., 2006, A\&A, 445, 29

\reference{} Prantzos, N.\ 2000, New Astronomy Review, 44, 303 

\reference{}Prantzos, N., \& Boissier, S., 2000, MNRAS, 313, 338 

\reference{}Renzini, A., 2006, ARA\&A, 44, 141


\reference{}Sabatini, S., Davies, J., van Driel, W., Baes, M., Roberts, S., Smith, R., Linder, S., O'Neil, K., 2005, MNRAS, 357, 819

\reference{}Samland, M., Hensler, G., \& Theis, C., 1997, ApJ, 476, 544 

\reference{}Sandage, A., Binggeli, B., Tammann, G., 1985, AJ, 90, 1759

\reference{}Scodeggio, M., Gavazzi, G., Franzetti, P., Boselli, A., Zibetti, S., Pierini, D., 2002, A\&A, 384, 812

\reference{}Sellwood, J., \& Carlberg, R., 1984, ApJ, 282, 61

\reference{}Seth, A., Dalcanton, J., de Jong, R., 2005, AJ, 130, 1574

\reference{}Silich, S., Tenorio-Tagle, G., 1998, MNRAS, 299, 249

\reference{}Silich, S., Tenorio-Tagle, G., 2001, ApJ, 552, 91

\reference{}Simien, F., Prugniel, P., 1997a, A\&AS, 122, 521

\reference{}Simien, F., Prugniel, P., 1997b, A\&AS, 126, 15

\reference{}Simien, F., Prugniel, P., 1998, A\&AS, 131, 287 

\reference{}Simien, F., Prugniel, P., 2002, A\&A, 384, 371

\reference{}Smail, I., Edge, A., Ellis, R., Blandford, R., 1998, MNRAS, 293, 124

\reference{}Somerville, R., Primack, J., 1999, MNRAS, 310, 1087


\reference{}Strader, J., Brodie, J., Spitler, L., Beasley, M., 2006, AJ, 132, 2333


\reference{}Tantalo, R., Chiosi, C., 2004, MNRAS, 353, 917 

\reference{}Thomas, D., Maraston, C., Korn, A., 2004, MNRAS, 351, L19

\reference{}Thomas, D., Brimioulle, F., Bender, R., Hopp, U., Greggio, L., Maraston, C., Saglia, R., 2006, A\&A, 445, L19

\reference{}Tinsley, B.~M.\ 1980,  Fundamentals of Cosmic Physics, 5, 287 

\reference{}Treu, T., 2004, in "Clusters of Galaxies: Probes of Cosmological Structure and Galaxy Evolution", Cambridge University Press,
ed. by Mulchaey et al., 2004, p. 178

\reference{}Treu, T., Ellis, R.~S., Kneib, J.-P., Dressler, A., Smail, I., Czoske, O., Oemler, A., \&  Natarajan, P.\ 2003, ApJ, 591, 53 

\reference{}Tully, B., Mould, J., Aaronson, M., 1982, ApJ, 257, 527

\reference{}Vader, J., 1986, ApJ, 305, 669
 

\reference{}van Zee, L., Barton, E., Skillman, E., 2004a, AJ, 128, 2797

\reference{}van Zee, L., Skillman, E., Haynes, M., 2004b, AJ, 128, 121

\reference{}V{\'a}zquez, G.~A., \& Leitherer, C.\ 2005, ApJ, 621, 695 

\reference{}Vilchez, J.M., 1995, AJ, 110, 1090

\reference{}Visvanathan, N., Sandage, A., 1977, ApJ, 216, 214
 



\reference{}Vollmer, B., Cayatte, V., Balkowski, C. \& Duschl, W., 2001, ApJ, 561, 708


%



\reference{}White, S., Rees, M., 1978, MNRAS, 183, 341

\reference{}White, S., Frenk, C., 1991, ApJ, 379, 52


\reference{}Yoshii, Y., Arimoto, N., 1987, A\&A, 188, 13


\reference{}Zwicky F., Herzog E., Karpowicz M., Kowal C., Wild P., 1968, 
"Catalogue of Galaxies and of Cluster of Galaxies" (Pasadena, California 
Institute of Technology; CGCG)

\end{document}